\documentclass[pra,twocolumn,showpacs,amsmath,amssymb]{revtex4}
\usepackage{epsfig}
\usepackage{color}
\usepackage{graphicx}
\usepackage{bm}
\usepackage{mathtools}

\DeclareMathOperator*{\osum}{\mathrlap{\hspace{0.8ex}\circ}{\sum}}

\usepackage[colorlinks,bookmarks=false,citecolor=blue,linkcolor=red,urlcolor=blue]{hyperref}

\begin{document}

 \title{Multiple Transitions of Coupled Atom-Molecule Bosonic Mixtures in Two Dimensions}

\author{L. de Forges de Parny$^1$, A. Ran\c{c}on$^1$, and T. Roscilde$^{1,2}$}  
\affiliation{$^1$ Laboratoire de Physique, CNRS UMR 5672,  \'Ecole Normale Sup\'erieure de Lyon, 
Universit\'e de Lyon, 46 All\'ee d'Italie, Lyon, F-69364, France}
\affiliation{$^2$ Institut Universitaire de France, 103 boulevard Saint-Michel, 75005 Paris, France}

\date{\today}

\begin{abstract}

Motivated by the physics of coherently coupled, ultracold atom-molecule mixtures, we investigate a classical model possessing the same symmetry --  namely a $U(1)\times \mathbb{Z}_2$ symmetry, associated with the mass conservation in the mixture ($U(1)$ symmetry), times the $\mathbb{Z}_2$ symmetry in the phase relationship between atoms and molecules. In two spatial dimensions the latter symmetry can lead to a finite-temperature Ising transition, associated with (quasi) phase locking between the atoms and the molecules. On the other hand, the $U(1)$ symmetry has an associated Berezinskii-Kosterlitz-Thouless (BKT) transition towards quasi-condensation of atoms or molecules. The existence of the two transitions is found to depend crucially on the population imbalance (or detuning) between atoms and molecules: when the molecules are majority in the system, their BKT quasi-condensation transition occurs at a higher temperature than that of the atoms; the latter has the  unconventional nature of an Ising (quasi) phase-locking transition, lacking a finite local order parameter below the critical temperature. When the balance is gradually biased towards the atoms, the two transitions merge together to leave out a unique BKT transition, at which both atoms and molecules acquire quasi-long-range correlations, but only atoms exhibit  conventional BKT criticality, with binding of vortex-antivortex pairs into short-range dipoles. The molecular vortex-antivortex excitations bind as well, but undergo a marked crossover from a high-temperature regime in which they are weakly bound, to a low-temperature regime of strong binding, reminiscent of their transition in the absence of atom-molecule coupling.

\end{abstract}


\maketitle

\section{Introduction}

 Ultra-cold atoms offer an entirely new approach to quantum many-body systems. One well-defined perspective is that of the controlled experimental implementation (or Feynman's analog quantum simulation) of models relevant to other fields, such as condensed matter or high-energy physics \cite{Bloch_2008, Lewensteinbook,Zoharetal2015}; or -- even more interestingly --  as quantum simulators of models which are not easily realizable in any other physical contexts, extending greatly our understanding and control over quantum many-body physics. One such example is represented by the possibility of coherently coupling different internal states of a particle (or of ensembles of particles): the states coupled together may have different physical properties, such as a different effective mass in a lattice, or different interparticle interactions, and hence can be effectively thought of characteristics of different physical particles. Such a possibility of coherent conversion among different particle species is rarely offered in physics -- similar examples being neutrino oscillations \cite{Suekanebook} or exciton-polariton condensates \cite{Sanvittobook}. Going beyond single-particle transmutation, cold atoms offer the possibility of coherently coupling pairs of atoms into molecules -- realizing in a sense quantum-coherent chemical reactions -- via the use of Feshbach resonances \cite{Chinetal2010} or photo-association \cite{Jonesetal2006}.  The coherent coupling is exquisitely quantum-mechanical, as it couples \emph{\`a la} Josephson the phase of the wavefunction of atom pairs with that of the molecule. When dealing with bosonic atoms and molecules, long-range phase coherence can be established at low temperature via Bose-Einstein condensation. In this context, the coherent coupling between atoms and molecules clearly leads to a complex interplay between the phenomena of condensation of atoms and molecules, and to the possibility of condensation transitions which are not realized in the context of mono-atomic quantum fluids. 
  
   In this paper we focus our attention on the intriguing case of two-dimensional atom-molecule mixtures. For these systems, Bose-Einstein condensation at finite temperature is impossible in the proper sense, leaving space to quasi-condensation via a Berezinskii-Kosterlitz-Thouless (BKT) transition \cite{Josebook}, associated with the unbinding transition of pairs of topological excitations (vortices and antivortices). In this context the coherent coupling between atoms and molecules is all the more interesting, as it establishes a correlation among the topological defects in the phase patterns of both species. Our main focus is on the temperature dependence of phase correlations in atom-molecule mixtures in the weakly interacting regime; in this perspective we can capture the universal features of the thermal phase transitions of the system by resorting to a simple model which assumes nearly homogeneous densities for both species, and which focuses uniquely on the local phases of the atomic and molecular field operators, treated as classical variables. The atomic and molecular phases are asymmetrically coupled in a double XY model, which has been very little investigated in the past. Here we provide an extensive Monte Carlo investigation of its critical properties, based on an original extension of the powerful Wolff cluster algorithm \cite{Wolff_1989} to the case of asymmetric XY interactions. The cluster algorithm gives access to the simulation of unprecedented system sizes: their analysis turns out to be crucial in order to critically revisit some existing claims of unconventional criticality \cite{Ghanbari_2005} and finite (local) Ising order \cite{Ejima_2011} in the model under investigation and closely related models.
   
   We reconstruct the phase diagram in the different regimes of strong vs. weak imbalance between atomic and molecular states, and of strong vs. weak atom-molecule coupling. In particular we unveil a complex interplay between the atom-molecule imbalance and coupling. Far on the molecular side of the resonance, the atom-molecule mixture is found to exhibit two transitions: a high-temperature BKT transition with onset of quasi-condensation for the molecules, and a low-temperature Ising transition for the quasi-condensation of atoms. Moving towards the atomic side, on the other hand, the two transitions merge into a unique BKT transition, with conventional quasi-condensation of the atoms and rather unconventional one for molecules, as the latter is fully driven by the coupling of molecules to atom \emph{pairs} and not to single atoms. The quasi-condensate phase is further characterized by a marked crossover between weakly bound and tightly bound molecular vortex-antivortex pairs. These rich phenomena are amenable to experimental verification using state-of-the-art setups in cold-atom physics. 
   
The paper is organized as follows. 
In Sec.~\ref{sec_models}, we review the microscopic model for atom-molecule mixtures, and connect it to the double XY model which will be the subject of the subsequent investigation.   
We discuss briefly the connection between the experimental control parameters and the microscopic parameters of the model; we describe its symmetries and we introduce the phases expected to appear. We also comment on the peculiarities of the two-dimensional case, and on the impossibility of a conventional breaking of the Ising symmetry at low-temperature in this model.
In Sec.~\ref{section3}, we discuss the Monte Carlo approach, based on a newly developed, generalized cluster algorithm, and we define the main observables of interest.
Sec.~\ref{sec_phasediag} is devoted to the discussion of the phase diagram. We discuss in further detail the molecular, 
atomic and resonant regime respectively in
Sections~\ref{sec_JallJm}, \ref{sec_JaggJm}  and \ref{sec_JasimeqJm}.  
Conclusions and outlook are provided in Sec.~\ref{conclusion}.

\section{Model Hamiltonians, and their theoretical and experimental relevance}
\label{sec_models}

\subsection{Hamiltonian for a resonant atom-molecule mixture}

A two-dimensional bosonic mixture of atoms and molecules, coherently coupled close to a resonance, can be described via the coupled-channel Hamiltonian
$\mathcal{\hat H}=   \mathcal{ \hat  H}_a+ \mathcal{\hat   H}_m+  \mathcal{\hat C}$, whose different terms read:
\begin{equation}
\begin{split}
& \mathcal{ \hat  H}_\alpha  = \int d^2 r \left\{\hat\psi_\alpha^{\dag}({\bm r})\left(-\frac{\hbar^2\nabla^2}{2 m_\alpha}-\mu_\alpha\right)\hat\psi_\alpha^{\phantom\dag}({\bm r})+\frac{g_\alpha}{2}\hat{\rho}_\alpha({\bm r})^2\right\} , \\
& \mathcal{\hat C}=\int d^2 r  \left\{g_{am}\hat{\rho}_a({\bm r})\hat{\rho}_m({\bm r})-\Gamma\left(\hat\psi_a^{\dag }({\bm r})^2\hat\psi_m^{\phantom\dag}({\bm r})+{\rm h.c.})\right)\right\} ~.
\end{split}
\label{eq_qH}
\end{equation}
Here  $\hat\psi_\alpha^{\dag}({\bm r})$ and $\hat\psi_\alpha^{\phantom\dag}({\bm r})$, ($\alpha=a, m$) 
are bosonic creation and annihilation operators of atoms and molecules respectively at position ${\bm r}$;
$\hat{\rho}_\alpha({\bm r})= \hat\psi_\alpha^{\dag}({\bm r})\hat\psi_\alpha^{\phantom\dag}({\bm r})$ is the corresponding density operator. 
$ \mathcal{ \hat  H}_\alpha$ corresponds to the Hamiltonian of a single species with mass $m_\alpha$, chemical potential $\mu_\alpha$, and intra-species interaction $g_\alpha>0$. 
The inter-species interactions are all included into $ \mathcal{\hat C}$, where the first term, proportional to $g_{am}>0$ corresponds to the repulsive inter-species interaction; whereas the second term, proportional to $\Gamma$, is the conversion term, that coherently transforms a pair of atoms into a molecule and vice versa \cite{Radzihovskyetal2008}. In the case of a magnetic Feshbach resonance \cite{Chinetal2010}, the difference between atomic and molecular chemical potentials can be changed experimentally because of the different Zeeman energies of atoms and molecules in the presence of a magnetic field; the field can be tuned to a Feshbach resonance \cite{Chinetal2010}, at which the atomic and molecular state come to degeneracy. As we will discuss later, the conversion can also be achieved with photo-association, in which case the chemical-potential difference between atoms and molecules is controlled by the sign of the detuning of the laser beams. 

The molecular state (closed channel) is usually eliminated adiabatically, and absorbed in a field-dependent scattering length for the atomic state (open channel). On the other hand, in the following we will consider both states as simultaneously present in the system, and focus particularly on the coherent coupling among them. 

On the theory side, the Hamiltonian of Eq.~\eqref{eq_qH} has been mainly the subject of mean-field studies \cite{Radzihovskyetal2004, Romansetal2004}, see Ref.~\onlinecite{Radzihovskyetal2008} for an extensive review. To go beyond the mean-field approximation a lattice version of the above model is usually introduced, motivated \emph{e.g.} by the possible application of an optical lattice. Most studies of coherently coupled atom-molecule mixture in a single-band lattice have focused on quantum phase transitions exhibited by the model in one dimension \cite{RousseauD2008, EckholtR2010, Ejima_2011, Bhaseenetal2012}  as well as in two dimensions \cite{SenguptaD2005, deForges_Roscilde_2015}. A detailed discussion of the experimental relevance of the model in Eq.~\eqref{eq_qH} is postponed to Sec.~\ref{s.exp}.

\subsection{Double XY model}

The mean-field approach \cite{Radzihovskyetal2008} for the Hamiltonian of the atom-molecule resonant mixture, Eq.~\eqref{eq_qH}, fails completely in the case of two spatial dimensions (2$d$)  at finite temperature. Indeed, due to the special role played by the long-wavelength fluctuations in 2$d$, the finite-temperature physics and critical phenomena are qualitatively and quantitatively different from the mean-field prediction. A theoretically sound treatment requires to take into account the full many-body physics of Eq.~\eqref{eq_qH}: this is a rather hard task, given the large number of independent parameters ($m_a, m_m, \mu_a, \mu_m, g_a, g_m, g_{am}, \Gamma$), and the need for a fully fledged numerical approach. Quantum Monte Carlo is in principle a viable approach to the case of bosonic gases, and it has been indeed used to study a lattice version of the atom-molecule Hamiltonian  \cite{RousseauD2008, deForges_Roscilde_2015}. On the other hand, if interested uniquely in the main, universal features of the finite-temperature phase diagram of the model, one can largely simplify the description of the system by resorting to a lattice classical model. Indeed, when studying finite-temperature transitions, a classical approximation $\hat\psi_{\alpha} \to \psi_{\alpha} = \sqrt{\rho_{\alpha}} e^{i\phi^{\alpha}}$ retains all the finite-$T$ universal physics. The only fundamental assumption behind this approximation is that the ground-state is a simultaneous condensate of both atoms and molecules, as the classical field theory will always predict this outcome. Moreover, as the critical phenomena in the system are fundamentally driven by phase fluctuations, while density fluctuations are not singular at the transition, one can neglect density fluctuations altogether, and focus uniquely on the phase degrees of freedom. In order to do numerics, it is very convenient to resort to a lattice version of the model, and this last step leads to the following double XY model 

\begin{eqnarray}
 \nonumber 
\mathcal{H}_{\rm XY}  =   &-& J_a \sum_{\langle {\bm r}, {\bm r'} \rangle}  \cos(\phi^a_{\bm r} - \phi^a_{\bm r'})   \\
  \nonumber 
  &-& J_m \sum_{\langle {\bm r}, {\bm r'} \rangle}  \cos(\phi^m_{\bm r} - \phi^m_{\bm r'})   \\
  &-& C   \sum_{{\bm r}}  \cos(\phi^m_{\bm r} - 2\, \phi^a_{\bm r})~.
\label{XY_hamiltonian}
\end{eqnarray}
Here $\langle {\bm r}, {\bm r'} \rangle$ are the bonds on any periodic 2$d$ lattice - and we choose the square lattice for definiteness. $J_a$ and $J_m$ are the characteristic energy scales for the kinetic energy of the atoms and molecules respectively, while $C$ is the characteristic atom-molecule conversion energy.  
The magnitudes of the effective parameters of the XY Hamiltonian can be readily extracted from the quantum many-body Hamiltonian of 
Eq.~\eqref{eq_qH}. Indeed, using elementary dimensional analysis one can estimate that, in a 2$d$ system
\begin{equation}
\begin{split}
J_\alpha&\sim  \frac{\hbar^2\bar{\rho}_\alpha}{2m_\alpha}\  ~~~~~~~ (\alpha=a,m),\\
C&\sim \Gamma (\bar{\rho}^2_a\bar{\rho}_m)^{\frac{1}{6}}~. 
\end{split}
\end{equation}
Here $\bar{\rho}_a$ and $\bar{\rho}_m$ are the average atomic and molecular densities, controlled by the chemical potentials $\mu_a$ and $\mu_m$, as well as by the density-density coupling constants $g_a$, $g_m$ and $g_{am}$. All these parameters do not appear explicitly in the XY Hamiltonian, but they enter indirectly via the densities. In particular the experimentally tunable atom-molecule detuning $\mu_a - 2\mu_m$ controls the ratio between the kinetic energies $J_a/J_m$, while the width of the Feshbach resonance, or the Rabi frequency of the effective atom-light coupling in the photo-association scheme, controls the coupling energy $C$ independently of the densities. Hence in the following we will assume that experiments can independently control the ratios $j = J_a/J_m$, $c = C/J_m$, and $t = T/J_m$, which naturally parametrize the phase diagram of the effective Hamiltonian in Eq.~\eqref{XY_hamiltonian}.  

\subsection{Hamiltonian symmetries, and expected phases at the mean-field level}
\label{sec_symmetries}

The atom-molecule coupling term in Eq.~\eqref{XY_hamiltonian} introduces a fundamental asymmetry between atoms and molecules in terms of the phase variable, and it dictates the peculiar symmetries of the Hamiltonian under phase transformations. 
When molecules and atoms are not coherently coupled ($C$$=$$0$), the model possesses two global $U(1)$ symmetries associated with the independent rotation of the phases of the fields, {\it i.e.} $\phi_{\bm r}^a\to\phi_{\bm r}^a+\theta_a$ and $\phi_{\bm r}^m\to\phi_{\bm r}^m+\theta_m$. For $C$$\neq$$0$, one cannot shift the two fields independently, and the $U(1)$$\times$$U(1)$ symmetry breaks down into a global  $U(1)$$\times$$\mathbb{Z}_2$ symmetry, corresponding to the transformations 
\begin{eqnarray}
\phi_{\bm r}^m&\to&\phi_{\bm r}^m+\theta \nonumber \\
\phi_{\bm r}^a &\to& \phi_{\bm r}^a+\frac{\theta}{2}+\frac{1}{2}(\sigma+1)\,\pi,  
\end{eqnarray}
where $\sigma=\pm 1$. Due to the atom-molecule phase coupling, the surviving $U(1)$ symmetry is a joint one for atomic and molecular phases, and indeed it corresponds to ``mass" conservation in the quantum Hamiltonian, namely the conservation of the quantity $\rho=\sum_r (\rho^a_r+2\rho^m_r)$.
If the atom-molecule coupling were of the standard Josephson type, namely of the kind $\cos(\phi^a - \phi^m)$, no further symmetry would be present -- this is the case \emph{e.g.} of the (one-body) Rabi coupling between different internal states, which has been the subject of several recent studies \cite{Niklasetal2011,Sartorietal2015}. 
 On the other hand, due to the asymmetric nature of the atom-molecule coupling, the model possesses a further $\mathbb{Z}_2$ symmetry which pertains uniquely to the atoms: as we shall discuss later, this emergent Ising symmetry is crucial for the understanding of the phase diagram.

Here we address the description of the different phases expected in the system at the level of the mean-field approximation \cite{Radzihovskyetal2008}. At high temperature, the system is disordered, corresponding to the normal or paramagnetic phase in the cold atoms and magnetic language respectively. As the temperature is lowered, there are two scenarios for the condensation or ordering transition, depending on the $j$ ratio. When $j\gg 1$ the atoms will drive the ordering process, condensing at a temperature $t \sim j$. Within the mean-field picture, the average phase of the atoms acquires a finite value, hence $\langle e^{i\phi^a} \rangle \neq 0$ and $\langle e^{i2\phi^a} \rangle \neq 0$. This in turns implies that the molecules condense at the same temperature, since their phase is locked to the non-zero value acquired by the phase of atomic pairs. The system is thus a joint molecular and atomic superfluid condensate (${\rm SF_{am}}$). 

On the other hand, when $j \ll 1$, the molecules condense first as the temperature is lowered towards $t \sim 1$, but this does not imply that atoms also order. Indeed, a finite value for the average $\langle e^{i \phi^m}\rangle$ couples to twice the phase of the atoms, which will then exhibit long-range order, namely $\langle e^{i 2 \phi^a}\rangle \neq 0$, but this fixes the phase of the atoms only modulo $\pi$. Therefore the molecular condensation leaves out the $\mathbb{Z}_2$ symmetry, which is only broken at a lower temperature $t \sim j$.  Thus, we expect that for $j \ll t \ll 1$ a molecular superfluid (${\rm SF_{m}}$) coexist with a normal atomic gas. For a temperature $t \sim j$, the atoms will also condense by choosing a definite phase between its two possibilities $ \phi^m/2+\frac{1}{2} (\sigma+1)\,\pi$ with $\sigma = \pm 1$. Hence the atom condensation is driven by the ordering of the Ising variable $\sigma$, corresponding to phase locking between atoms and molecules, and we can thus expect the physics of this transition to be described by an effective Ising model, as we will further elucidate below.

\subsection{Why $2d$ is special: is $\mathbb{Z}_2$ really broken at finite temperature? \label{sec_z2}}

\subsubsection{Absence of a finite local Ising order parameter}

As already mentioned, fluctuations have dramatic effects in 2$d$, drastically altering the mean-field picture. In particular, true long-range order (LRO) or condensation are impossible \cite{Mermin}, and they are replaced by quasi-LRO or quasi-condensation \cite{Josebook}, characterized by algebraically decaying phase correlations. Nonetheless, as we will further elaborate in the following, for $j \gg 1$ the onset of quasi-LRO in the atoms drives necessarily the appearance of quasi-LRO in the molecules, albeit with a fundamental difference in the decay exponent of correlations. On the other hand, for $j \ll 1$ the onset of quasi-LRO in the molecules does not entail the same phenomenon in atoms, and the quasi-condensation transition of atoms remains decoupled from that of the molecules, retaining its peculiar Ising nature. To flesh out the Ising symmetry associated with the atoms, one can use the following decomposition for the atomic phase:  
 \begin{equation}
 \phi_{\bm r}^a = \tilde\phi_{\bm r}^a + \frac{1}{2} (\sigma_{\bm r} +1) \pi
 \label{eq_phidec}
 \end{equation}
 where $\tilde\phi_{\bm r}^a \in [0,\pi]$ is a reduced phase variable which describes the direction of the atomic spin, whereas $\sigma_{\bm r}=\pm 1$ gives its orientation.
 The atom-molecule coupling only involves the reduced phase variable, as $\cos(2\phi^a-\phi^m) =  \cos(2\tilde\phi^a-\phi^m)$, while it leaves the $\sigma$ variable free to fluctuate. Hence, one is immediately tempted to identify the Ising transition of the system with the onset of long-range order in the latter variable.
  
 Yet, concerning the ordering of $\sigma_{\bm r}$ variable at finite temperature, a fundamental remark is in order (see also \cite{Korshunov1985}). 
Given that $\tilde\phi^a$ is not a true angular variable but a reduced one, restricted to $[0,\pi]$, one has that
\begin{equation}
\langle \exp(i\tilde{\phi}^a) \rangle \neq 0
\end{equation}
 under \emph{all} physical circumstances (namely even at extremely high temperatures). This leads to the following, drastic consequence: 
 \begin{equation}
 \langle \sigma \rangle \neq 0 \to \langle \exp(i\phi^a) \rangle = \langle - \sigma  \exp(i\tilde{\phi}^a) \rangle \neq 0
\end{equation}
namely true long-range Ising order implies automatically atomic condensation. The latter is impossible in 2$d$, which in turn implies necessarily that $\langle \sigma \rangle = 0$ at any finite temperature. This means that the $\mathbb{Z}_2$ symmetry associated with the $\sigma$ variable remains effectively unbroken, namely the $\sigma$ variable can develop \emph{at most} algebraic correlations. 
This implies that only \emph{quasi} phase locking is realized between atoms and molecules, in the same sense that atoms and molecules only quasi-order at low temperature.
This is not in contradiction with the existence of an Ising critical point at a critical temperature $T=T_c$, at which $\langle \sigma_{\bm r} \sigma_{\bm r'} \rangle \sim |{\bm r} - {\bm r}'|^{-\eta_\sigma(T_c)}$ with $\eta_\sigma(T_c) = 1/4$ -- and indeed our data are fully consistent with the existence of such a critical point. Yet, below this critical point, the correlations of the $\sigma$ variable must necessarily remain algebraic with an exponent $\eta_\sigma(T)$ which continuously goes to zero as $T\to 0$, similar in spirit to what happens to spin-spin correlations in the XY model. The `magnetization'  $\langle \sigma \rangle$ is expected to vanish as $N^{-\eta_\sigma(T)}$ in the thermodynamic limit ($N$ is the number of spins).

\subsubsection{Finite-size effects: from the size of Texas to the size of the Universe (and beyond)}

To support the above argument (namely the absence of true atomic condensation in 2$d$), we will estimate a finite $\eta_\sigma(T)$ exponent for $T\leq T_c$ which decreases to zero very rapidly with temperature, as shown in Sec.~\ref{s.Ising}.   In the face of such strong finite-size effects, one can legitimately ask if the absence of Ising long-range order is actually observable experimentally. It is well known that finite-size effects are strongly relevant in the standard XY model, where it has been famously argued \cite{Texas} that in order to observe a magnetization of order $10^{-2}$ at the critical point, one would need a system of the size of the state of Texas (assuming a distance of 1 \AA ~between the microscopic degrees of freedom, as in a solid -- a much larger size would be required in diluted systems such as cold-atom samples).
 A similar argument shows that, already for $T=0.99~ T_c$, one would need a system size several orders of magnitude larger than that of our Universe in order to observe an Ising magnetization of $10^{-2}$.  Hence effective long-range Ising order will manifest itself in any experimental realization.
This aspect might also be at the origin of the conclusions of Ref.~\onlinecite{Ejima_2011}, claiming the numerical observation of long-range order of the Ising variable for a quantum $(1+1)$-dimensional realization of the transition that we discuss in this work. Ref.~\onlinecite{Ejima_2011} seemingly overlooks the fact that long-range order of the Ising variable entails automatically atomic condensation in 1$d$, hence its conclusions need to be reconsidered.

\subsubsection{From local to global $\mathbb{Z}_2$ order parameter?}

 Given that the Ising variable $\sigma$ cannot display true long-range order at finite temperature, it is legitimate to ask whether the transition for the onset of atomic quasi-condensation in 2$d$ is really accompanied by the breaking of some (hidden) $\mathbb{Z}_2$ symmetry. 
 A putative Ising variable developing long-range order cannot be a strictly local one (such as $\sigma$), as it would necessarily entail the appearance of atomic condensation -- as already noticed in Ref.~\onlinecite{Korshunov1985}. 
 Yet, in the limit $c\to\infty$ (or equivalently, in the context of the $\varphi/2\varphi$ model discussed below), 
a mapping to the Coulomb gas shows that non-local Ising variables can be introduced  \cite{Korshunov1985, Lee1985},
which nonetheless display long-range order in the ${\rm SF_{m}}$ phase while they are disordered in the ${\rm SF_{am}}$ -- 
a behavior which is dual to the one expected for an Ising variable ordering at the transition. 
The problem of identifying long-range Ising order in the ${\rm SF_{am}}$ phase (albeit of non-local nature) remains therefore open.

\subsection{Previous studies and related models}

Previous studies addressed the model of Eq.~\eqref{XY_hamiltonian}, and closely related models.
In the context of smectic liquid crystals, this model has been studied in 2$d$ by means of Monte Carlo simulations in Ref.~\onlinecite{Shahbazi_2006}. The latter study focuses on the regime $j\sim 1$, in which both atomic and molecular quasi-LRO establish simultaneously, and it claims that non-universal, $j$-dependent critical behavior is observed at the transition. In the following we will disprove this claim, showing that it is misled by the small system sizes explored in Ref.~\onlinecite{Shahbazi_2006}, while the larger system sizes that we can access reconcile all results with a universal scenario for the transition.
Motivated by the physics of liquid crystals, several numerical \cite{Jiang_1993, Jiang_1996, Ghanbari_2005} and renormalization group  \cite{Granato_1986, erratum_Granato, Bruinsma_1982, Kohandel_2003} studies have been devoted to an Hamiltonian of the form Eq.~\eqref{XY_hamiltonian}, but with a generalized coupling term of the form $\cos(p\phi^a -\phi^m)$ with integer $p$. This would correspond, in the cold-atom context, to a coherent atom/$p$-mer coupling.
 
  Finally, the $C$$\to$$\infty$ limit locks rigidly the atomic and molecular phases, allowing one to eliminate one of the two variables. Eliminating the molecular variable with the substitution $\phi^m = 2\phi^a$, one immediately resorts to the \emph{$\varphi/2\varphi$ model}  
  \begin{equation}
  H  =   - \sum_{\langle r,r' \rangle} \left [  \Delta  \cos(\varphi_{\bm r} - \varphi_{\bm r'}) + (1-\Delta)  \cos(2\, \varphi_{\bm r} - 2\, \varphi_{\bm r'}) \right ]
  \label{e.delta2delta}
  \end{equation}
  with $\varphi = \phi^a$ and $\Delta = j/(j+1)$ (up to a global multiplicative constant for the Hamiltonian). This model has been widely investigated in the past \cite{Korshunov1985, Lee1985,CarpenterC1989,Shi2011,HubscherWessel_2013}, especially in connection with the physics of bosonic pairing \cite{BonnesW2011}. Since the adiabatic elimination of the molecular field operator in Eq.~\eqref{eq_qH} would produce a pairing term in the quantum Hamiltonian, it is intuitive that the phase diagram of Eq.~\eqref{e.delta2delta} has quite some relevance to the model of our interest. 

\subsection{Experimental relevance} 
\label{s.exp}

 As already mentioned in the introduction, a natural realization of the atom-molecule Hamiltonian of Eq.~\eqref{eq_qH} is in the context of ultracold bosons close to a magnetic Feshbach resonance.  Feshbach resonances have been widely used to tune the scattering length of bosonic gases \cite{Chinetal2010}, controlling in this way the coherence properties of atomic Bose-Einstein condensates (BECs). Recently, significant attention has been devoted to the unitary Bose gas with infinite scattering length \cite{Remetal2013, Fletcheretal2013, Makotynetal2014}, subject to rapid decay due to three-body recombination of atoms and inelastic collisions of molecules. More relevant to this work, the coherent coupling $\hat{\cal C}$ (in the form of Rabi oscillations) between two unbound bosonic atoms and a weakly bound molecular state at a magnetic Feshbach resonance has been first demonstrated in Ref.~\onlinecite{Donleyetal2002}, and subsequently shown in other remarkable experiments starting either from a Bose-Einstein condensate \cite{Thompsonetal2005} or a Mott insulator in a deep optical lattice \cite{Syassenetal2007}. The association of Feshbach molecules starting from an atomic Bose-Einstein condensate suggests the possibility of creating an equally coherent molecular Bose-Einstein condensate 
\cite{Kokkelmansetal2001}. Yet this perspective is hindered by the weakly bound nature of the Feshbach molecules, which are collisionally unstable and subject to heating: this is due to processes which transfers the internal energy of a molecule (decaying to a deeply bound state) to internal and kinetic energy of another molecule, which typically breaks up into two thermal atoms \cite{note}.
Remarkably, a very high loss rate may in fact inhibit molecular collisions, leading to effective hardcore repulsion and stabilization of the molecular gas \cite{Syassenetal2008}. 

  Realizing a BEC of weakly bound Feshbach molecules out of an atomic BEC remains a technical challenge, unachieved so far. A valid alternative (or complement) to Feshbach association of molecules is represented by photo-association, which has the advantage of creating deeply bound molecules (down to their rotovibrational ground state \cite{Danzletal2008, Langetal2008}), and which has recently been demonstrated to lead to coherent oscillations between atomic and molecular BECs \cite{Yanetal2013}. 
In the following we will assume that the coherent atom-molecule coupling $\Gamma$ in Eq.~\eqref{eq_qH} can come from either hyperfine coupling at a magnetic Feshbach resonance, or that it is induced by lasers in a single-photon or two-photon scheme. In the latter case the control on the chemical-potential difference between atoms and molecules is provided by the laser detuning. 

 Despite the technical challenges offered by the creation of a stable molecular BEC starting from an atomic BEC, its achievement opens the route to a very rich phenomenology. A lot of attention has been devoted recently to the consequences of the formation of stable heteronuclear molecules, that may develop a large electric dipole under an applied electric field \cite{Lahayeetal2009, Baranovetal2012}. In this paper we would like to point out that the formation of a stable BEC of \emph{homonuclear} molecules, coherently coupled to an atomic BEC, can also open the path to very rich physics. Here we will specialize our attention to the rich phase diagram in the case of two-dimensional (quasi-)condensates; the phase diagram is as rich in three spatial dimensions, whose treatment we shall postpone to future work.

Finally, in the specific context of two-dimensional systems it is worth mentioning the possibility of realizing our model of interest in the context of microcavity polariton fluids. Such systems also enjoy Feshbach resonances \cite{Takemuraetal2014} and their molecular states (bipolaritons) may have a pronounced stability, leading to the observation of a molecular (quasi-)condensate \cite{MarchettiK2014}, so far investigated theoretically only in the case of polaritons in different spin states. 

\section{Cluster Monte Carlo approach and observables}
\label{section3}

The numerical treatment of critical phenomena, associated with a diverging correlation length, can be very efficiently tackled via the Monte Carlo approach when equipped with cluster algorithms \cite{Swendsen_Wang,Wolff_1989}, which beat the critical slowing down of local dynamics. If the original Wolff's cluster algorithm \cite{Wolff_1989} is readily applicable to conventional XY models (such as Eq.~\eqref{XY_hamiltonian} with $C=0$), in the case of coupled XY models of our interest this algorithm needs to be appropriately generalized. 
Our newly introduced algorithm (described in Appendix \ref{app_wolff}) is in fact able to generate a cluster update for two coupled fields with a generic coupling of the form $\cos(\phi^m - p\phi^a)$ with $p \in \mathbb{N}$, for which the conventional Wolff algorithm can no longer be applied as soon as $p>1$. Previous studies relied on simple single-flip Metropolis updates, or on a mix of Wolff updates (for the intra-layer interaction) and Metropolis updates (to treat the inter-layer interaction)  \cite{Jiang_1993,Jiang_1996,Ghanbari_2005,Shahbazi_2006} -- while the first scheme suffers from critical slowing down at the transition, the latter scheme suffers from an exponential rejection rate when the temperature is lowered. 

Our simulations are performed on a square lattice of size $L\times L$ ($L$ ranging from $10$ to $400$) with periodic boundary conditions. Hereafter we use the lattice spacing as unit length.
Our analysis of the phase diagram of the system is based on several thermodynamic quantities that we list below.

A generic signature of conventional phase transitions is provided by the specific heat  
\begin{eqnarray}
C_V= \frac{1}{L^2}\partial \langle E \rangle / \partial T  =  \frac{\langle E^2 \rangle -  \langle E \rangle^2}{L^2 T^2} ~,   
\label{Cv}
\end{eqnarray}
where $\langle E \rangle=\langle \mathcal{H}_{XY} \rangle$ is the total energy ($k_B$=1), $T$ the temperature, and 
$ \langle . \rangle \equiv \frac{1}{N_{\rm MC}} \sum_n^{N_{\rm MC}} (.)$ is the Monte Carlo average. The specific heat will be essential to detect Ising transitions, while it only provides a rounded maximum close to (but above) BKT transitions.  

On the other hand BKT transitions are signaled by the appearance of quasi-LRO, which is detected via a $k=0$ peak in the momentum distribution, which we shall define as follows \cite{caveat}
 for atoms (molecules):
\begin{eqnarray}
n^{a(m)}_0= \frac{1}{L^2} \left \langle  \sum_{{\bm r},{\bm r'}}  \cos\left(\phi^{a (m)}_{\bm r} - \phi^{a (m)}_{\bm r'}\right)  \right \rangle~.\   
\label{momentum_distribution}
\end{eqnarray}
This definition coincides with the static structure factor when thinking of planar spins instead of the phase of bosonic fields. 
In the following we will denote $n_0^{a(m)}$ as the atomic (molecular) quasi-condensate.  

 It will also be important to monitor the coherence of atom pairs, whose momentum distribution is estimated as 
\begin{eqnarray}
n^{aa}_0= \frac{1}{L^2} \left \langle  \sum_{{\bm r},{\bm r'}} \cos\left(2\,\phi^{a}_{\bm r} - 2 \,\phi^{a}_{\bm r'}\right)  \right \rangle~.\   
\label{pair_momentum_distribution}
\end{eqnarray}

Furthermore, to study the expected Ising transition,
it is convenient to define the (local)  Ising variable  $\sigma_{\bm r}=-\sin(\phi^{a}_{\bm r})/|\sin(\phi^{a}_{\bm r})|=\pm1$ and its related structure factor 
\begin{eqnarray}
n^\sigma_0=  \frac{1}{L^2} \left \langle  \sum_{{\bm r},{\bm r'}}    \sigma_{\bm r} \sigma_{\bm r'} \right \rangle~.
\label{Ising_momentum_distribution}
\end{eqnarray}

To access the superfluid response of the system, we calculate the  spin stiffness $\Upsilon$ (or helicity modulus),
 defined in terms of the second derivative of the free-energy
density  $f$ of the system with respect to a twist $\varphi$ along a given ($x$) direction, 
$\Upsilon=(\partial^2  f/ \partial \varphi^2)\arrowvert_{\varphi=0}$. The twist needs to be taken in a way that respects the \emph{joint} $U(1)$ symmetry of the coupled atomic and molecular fields; namely a twist $\varphi$ for the phase of the atomic field, $\phi_{\bm r}^{a} \to \phi_{\bm r}^{a} + \varphi ~{\bm r}\cdot \hat{x}$ must be accompanied by a twist of $2 \varphi$ for the molecules, $\phi_{\bm r}^{m} \to \phi_{\bm r}^{m} + 2\varphi ~{\bm r}\cdot \hat{x}$. In this way the atom-molecule conversion is unaffected by the twist. 
This leads to the following expression for the joint superfluid stiffness of the atom-molecule mixture 
\begin{eqnarray}
\nonumber
\Upsilon &=&  \frac{1}{L^2}  \left \langle  \sum_{ \langle {\bm r},{\bm r'} \rangle_x}  \left( J_a \cos( \phi^{a}_{{{\bm r}{\bm r}'}}) +4  J_m \cos( \phi^{m}_{{\bm r}{\bm r}'}) \right ) \right \rangle \\
&-&  \frac{1}{T L^2}  \left \langle  \left [ \sum_{ \langle {\bm r},{\bm r'} \rangle_x}  \left ( J_a \sin( \phi^{a}_{{\bm r}{\bm r}'}) +2 J_m \sin  ( \phi^{m}_{{\bm r}{\bm r}'}) \right) \right]^2 \right \rangle \ , \ \ \ \ \ \
\label{spin_stiffness}
\end{eqnarray}
with $\phi^{a(m)}_{{\bm r}{\bm r}'} = \phi^{a(m)}_{\bm r} - \phi^{a(m)}_{\bm r'}$.

Finally, the topological defects in the phase pattern are detected by studying the vorticity around the square plaquettes (${\Box}$) of the lattice. 
The local vorticity around a given plaquette ${\Box}$ is then defined as 
\begin{equation}
\rho_{V,\Box}^{a(m)}=   \frac{1}{2\pi} \osum_{{\bm r}\to{\bm r'}} \nabla \phi^{a(m)}_{{\bm r}{\bm r}'}  ~.
\label{vorticity}
\end{equation}
where the sum runs over the bonds of the plaquette oriented in a counterclockwise fashion, and the lattice gradient $\nabla \phi^{a(m)}_{{\bm r}{\bm r}'}$ corresponds to the phase difference $\phi^{a(m)}_{{\bm r}{\bm r}'}$ defined over the interval $[-\pi,\pi]$, namely it takes the value $\phi^{a(m)}_{{\bm r}{\bm r}'} - 2\pi  ~~( + 2\pi)$ if  $\phi^{a(m)}_{{\bm r}{\bm r}'} > \pi ~~ ( < -\pi)$.  

We also define a total vortex-antivortex pairs density, which only probes the vorticity amplitude at each plaquette, regardless of its sign
\begin{equation}
\rho_{V-AV}^{a(m)}=  \frac{1}{2 L^2}  \left \langle \sum_{\Box}  \left |  \rho_{V,\Box}^{a(m)}  \right | \right \rangle~.
\label{totalvorticity}
\end{equation}

\begin{figure*}
\begin{center}
\includegraphics[width=2 \columnwidth]{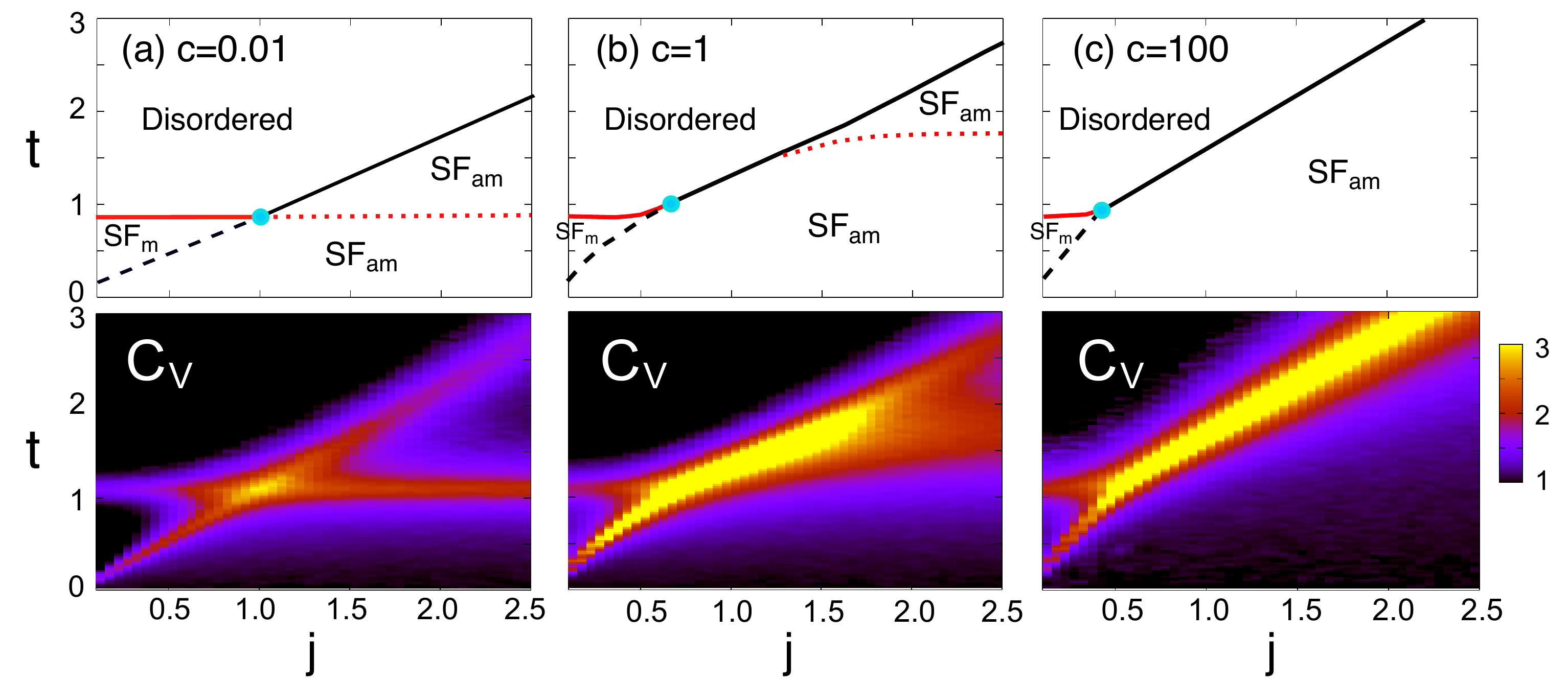}
\caption{ (Top row)  Phase diagram in the $(j,t)$ plane; (bottom row) specific heat $C_V$ for $L=20$  in false colors. 
The three columns refer to three values of the atom-molecule:  (a) small ($c=0.01$),  (b) moderate ($c=1$), and (c) large ($c=100$). 
Three phases are observed: a disordered phase at high temperature,  an atomic-molecular superfluid (${\rm SF_{am}}$) and a molecular superfluid (${\rm SF_{m}}$)  at low temperature. 
The black and red solid lines are molecular and atomic BKT transitions respectively. 
The black dashed line indicates a 2$d$ Ising transition whereas the red dotted line marks a crossover temperature  for the strong binding of molecular vortices (Sec.~\ref{sec_JaggJm}).
The light blue dot shows the probable location of the tricritical point.}
\label{diagramme_phase}
\end{center}
\end{figure*}

\section{Overview of the phase diagram}
\label{sec_phasediag}

The model Hamiltonian $\mathcal{H}_{\rm XY}$ Eq.~\eqref{XY_hamiltonian}  has a rich physics inherent to its two-dimensional nature.
To fix the ideas, let us consider the trivial case $C=0$, 
where both atomic and molecular fields admit a BKT transition 
at the critical temperatures $T^{a(m)}_{\rm BKT}/J_{a(m)} \simeq 0.89$ \cite{2DXY}, separating a high-temperature normal phase from a low-temperature superfluid (SF) quasi-condensate phase. This implies four different phases: normal, atomic SF (${\rm SF_{a}}$), molecular SF (${\rm SF_{m}}$) and joint atomic/molecular SF (${\rm SF_{am}}$), none of which breaks the $U(1) \times U(1)$ symmetry of the two decoupled bosonic fields. 

 Turning on the coupling $C$ completely changes the phase diagram, see Fig.~\ref{diagramme_phase}. The model exhibits three phases, which are the 2$d$ analogues of the phases discussed in Sec.~\ref{sec_symmetries} at the mean-field level: beside the normal phase for both atoms and molecules, we observe an ${\rm SF_{m}}$, with unbroken symmetries and quasi-LRO exhibited uniquely by the molecular field, and an ${\rm SF_{am}}$, exhibiting quasi-LRO for both fields.

 The succession of phases crucially depends on the ratio $j$ between atomic and molecular kinetic energies. When $j\ll 1$ (namely the \emph{molecular regime} of the model) the system exhibits two phase transitions: a higher BKT transition for the onset of the SF$_m$ phase, and a lower Ising transition marking the onset of the SF$_{am}$ phase. On the other hand, when $j\gg 1$ (or the \emph{atomic regime} of the model), there is a unique BKT transition, marking the onset of the SF$_{am}$, albeit with very different algebraic correlations for the atoms (which exhibit standard BKT phenomenology) and molecules (displaying unconventional algebraic correlations). When $c \ll 1$, the specific heat may erroneously suggest the existence of a second, lower-temperature transition, which could be naively associated with the onset of quasi-LRO in the molecules. In fact, as we shall later discuss, there is no separate transition for the molecules, and the binding of atomic vortex-antivortex (V-AV) pairs induces a similar binding phenomenon for the molecular V-AV excitations. What survives of the molecular transition for $c=0$ is just a crossover (marked with red dotted lines in Fig.~\ref{diagramme_phase}(a-b)) at which molecular V-AV pairs go from weakly bound to strongly bound. The crossover is sharper the lower $c$, while it essentially disappears when the BKT transition occurs at a temperature $t \lesssim c$.
 In the regime intermediate between the molecular and the atomic one ($j \sim 1$), which we shall later call the \emph{resonant regime}, we find that the two transitions of the molecular side merge into a unique transition,  which, similarly to the atomic regime, is driven by a standard atomic BKT transition.
 
 As already alluded at above, the strength of the atom-molecule coupling $c$ does not seem to change in any fundamental way the structure of the phase diagram, except for making more or less visible some selected features thereof. In particular, if $c \ll 1$, the energy scales of the Hamiltonian terms inducing coherence among atoms on one side, and molecules on the other, remain well separated in the opposite regimes $j \ll1$ and $j \gg 1$, so that one can clearly resolve the two transitions (molecular BKT transition and atomic Ising transition) in the molecular regime, and the separation between the joint BKT transition and the molecular crossover in the atomic regime. On the other hand, as it can be easily guessed, a large $c$ greatly enhances the resonant regime possessing a single characteristic temperature, and it pushes it far beyond the region with $j \sim 1$. 
 
 In the following sections we present compelling numerical evidence for this rich phenomenology in the three regimes identified above.

\begin{figure}[t!]
\begin{center}
\includegraphics[width=1 \columnwidth]{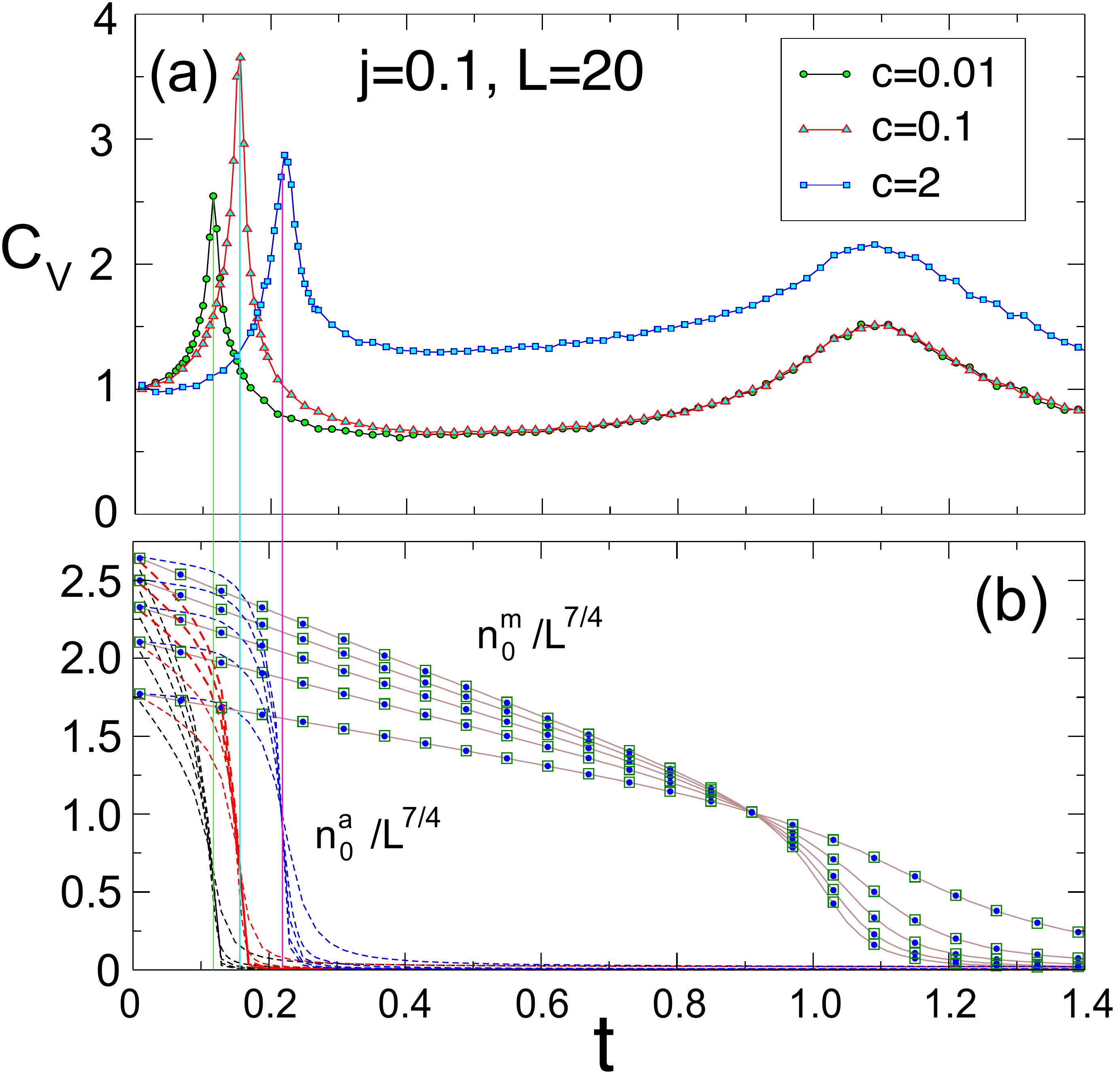}
\caption{Transitions in the atomic regime, $j=0.1$.
(a) Specific heat $C_V$ for $L$$=$$20$ and three values of the coupling constant, $c$$=$$0.01, 0.1$ and 2. The two maxima in $C_V$ are 
related to two successive transitions when increasing the temperature form zero: the Ising transition and then the BKT transition at $t^m_{BKT}$$\simeq$$0.89$.
(b) Finite size scaling of the atomic and molecular quasi-condensates $n^a_0$ and $n^m_0$ for $L$$=$$10, 20, 30, 40, 50$ 
($n^m_0$ is the same for these three $c$ values).
The atomic transition is a 2$d$ Ising transition whereas the molecular transition is a BKT transition.
}
\label{Cv_scaling_condensates_2DXY_Wolff_Ja1_Jm10}
\end{center}
\end{figure}  
 
\begin{figure}[t!]
\begin{center}
\includegraphics[width=1 \columnwidth]{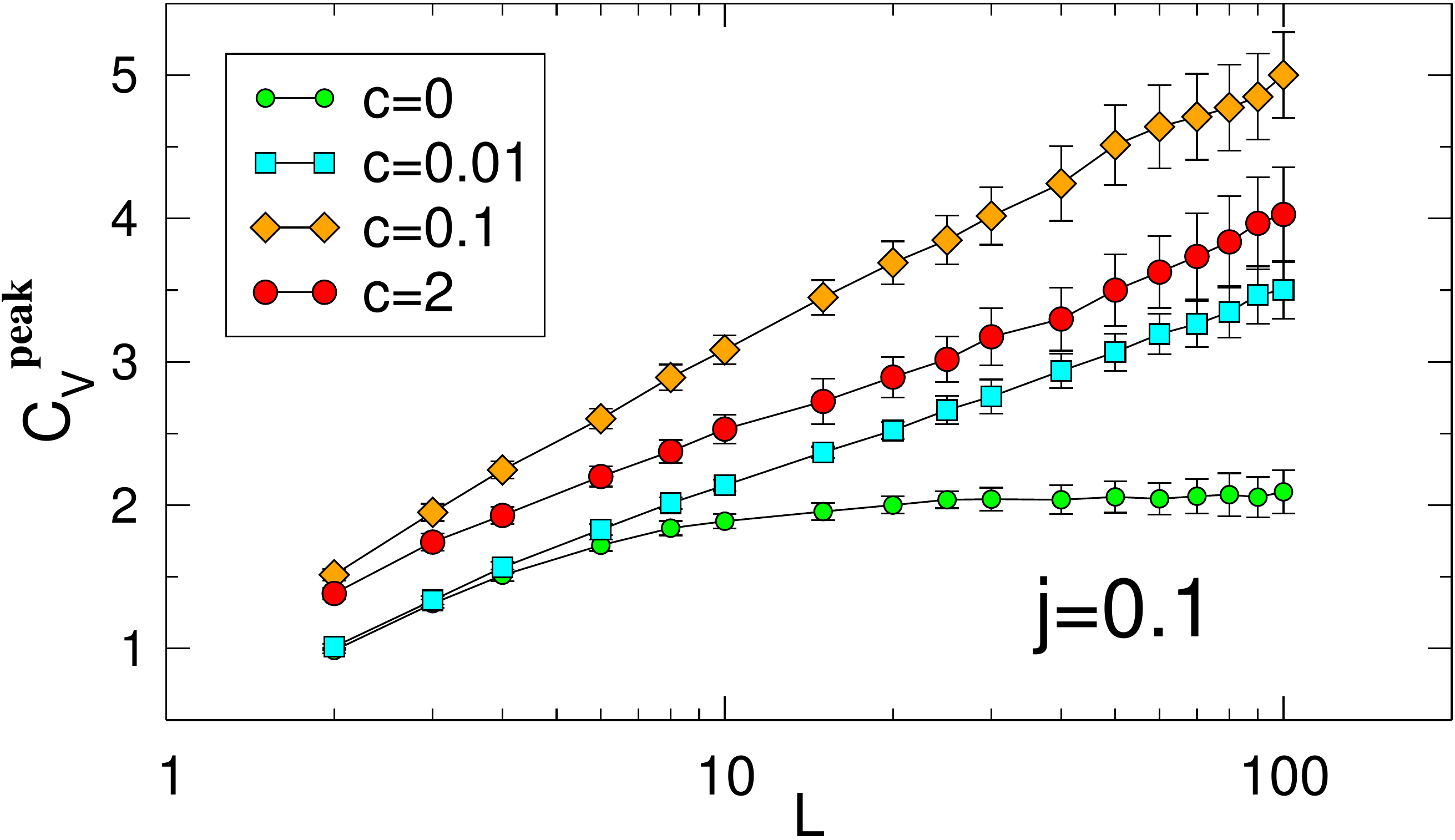}
\caption{ Scaling of the $C_V$ maximum  at the atomic transition $t^a_c$ versus $L$ in semi-log scale for $j=0.1$ and 
different coupling. For $c\neq0$, the specific heat scale as $C^{\rm peak}_V\propto \ln(L)$ in agreement with the 2$d$ Ising universality class (the exponent $\alpha=0$), even for
a very small coupling $c=0.01$. At $c=0$, the specific heat saturates, as expected for a BKT transition.}
\label{Scaling_Cv_Ising_2DXY_Wolff_Ja1_Jm10_Cvarie}
\end{center}
\end{figure}  

\begin{figure}[t!]
\begin{center}
\includegraphics[width=1 \columnwidth]{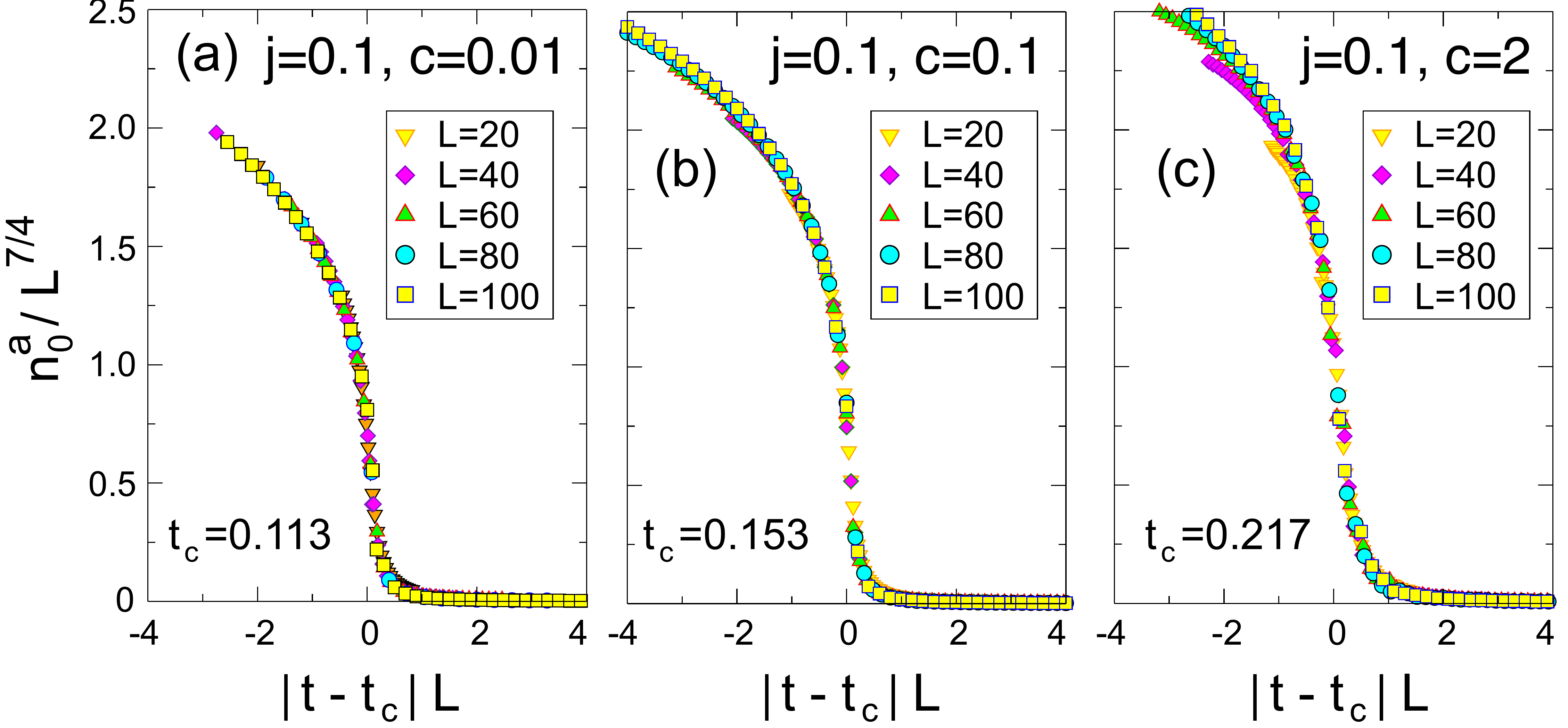}
\caption{ Collapse of the atomic quasi-condensate $n^a_0$ using 2$d$ Ising critical exponents ($\gamma$$=$$7/4$ and $\nu$$=$$1$)
for $j $$= $$0.1$ and various values of $c$.
}
\label{Scaling_total_na_Ja1_Jm10_CsurJm_0p01_0p1_2}
\end{center}
\end{figure}

\section{Molecular regime: $j \ll 1$ }
\label{sec_JallJm}

The molecular regime $j \ll 1$ is characterized by two peaks in the specific heat, as already seen in Fig.~\ref{diagramme_phase}. 
Fig.~\ref{Cv_scaling_condensates_2DXY_Wolff_Ja1_Jm10} (a) shows the two maxima for $j=0.1$ and various values of $c$: one may immediately notice that the low-$T$ maximum is a sharp one, while the high-$T$ maximum is a rounded one, consistently with the lower one being associated with an Ising transition and the higher one with a BKT transition. 
We first analyse the Ising transition, and then turn to an analysis of the BKT one. 

\subsection{Atomic Ising transition}
\label{s.Ising}

The putative Ising transition temperature is estimated via a scaling analysis of the atomic quasi-condensate (Fig.~\ref{Cv_scaling_condensates_2DXY_Wolff_Ja1_Jm10}(b)): the temperature at which $n^a_0\propto L^{\gamma/\nu}$, with $\gamma=7/4$ and $\nu=1$ is in very good agreement with the low-$T$ specific heat peak. Moreover the latter peak is shown in Fig.~\ref{Scaling_Cv_Ising_2DXY_Wolff_Ja1_Jm10_Cvarie} to scale logarithmically with system size when $c>0$, again consistently with a 2$d$-Ising transition, while at $c=0$ the peak saturates, as it corresponds to a BKT transition.

Further compelling evidence for the Ising transition is provided by the scaling plots of Fig.~\ref{Scaling_total_na_Ja1_Jm10_CsurJm_0p01_0p1_2} where the atomic quasi-condensate is found to obey the scaling the form
\begin{equation}
n^a_0 = L^{\gamma/\nu} f( |t-t_c| L^{1/\nu})
\label{eq_nascal}
\end{equation}
with critical exponents $\gamma$ and $\nu$ of the 2$d$ Ising model. This is rather remarkable, as it shows that quasi-LRO sets in obeying Ising criticality instead of BKT criticality  -- namely the quasi-condensate (corresponding to the susceptibility) and the correlation length for phase correlations diverges according to the Ising critical exponents. And this in spite of the fact that the low-temperature phase in the system is very different from that of an Ising model, as already discussed in Sec.~\ref{sec_z2} and further discussed below.  Note that Eq.~\eqref{eq_nascal} works for all values of the coupling constant $c$: in fact $c$ is a relevant perturbation in the renormalization group sense, and along the renormalization-group flow it locks the atomic and molecular phases up to the value of the Ising variable, namely $\phi^a_{\bm r}\simeq \phi^m_{\bm r}/2+ \frac{1}{2} (\sigma_{\bm r}+1) \pi $ close enough to the critical point (see also Sec.~\ref{sec_JaggJm} for further discussion).

The emergence of Ising physics can be further understood in the light of the decomposition of the atomic phase into a reduced phase variable and Ising variable as in Eq.~\eqref{eq_phidec}. 
As a consequence, the atomic correlation function decomposes as
 \begin{equation}
 C^a_{\bm r, \bm r'} = \langle \cos(\phi_{\bm r}^a-\phi_{{\bm r}'}^a) \rangle = \langle \sigma_{\bm r} \sigma_{{\bm r}'}  \times \cos(\tilde \phi_{\bm r}^a-\tilde \phi_{{\bm r}'}^a) \rangle ~.
 \label{e.corr_decomposition}
 \end{equation}
As discussed in Sec.~\ref{sec_z2}, the correlations of the reduced angular variable $\tilde{\phi}^a$ or $\phi^m/2$  are \emph{always} long-ranged, namely 
 \begin{eqnarray}
 \langle \cos(\tilde \phi_{\bm r}^a-\tilde \phi_{{\bm r}'}^a) \rangle & \to  & \langle \sin(\tilde\phi_{\bm r}^a) \rangle  \langle \sin(\tilde\phi_{\bm r'}^a) \rangle \neq 0 
  \end{eqnarray}
  for $|{\bf r}-{\bf r}'|\to \infty$, 
 where the average of the $\sin$ function does not vanish because of the $[0,\pi]$ range of the reduced angular variable $\tilde\phi^a$. This generically implies that the atomic correlations are \emph{always} dominated by the decaying correlations of the Ising part, namely $C^a_{\bm r, \bm r'}  \sim \langle \sigma_{\bm r} \sigma_{\bm r'} \rangle$.  Yet the Ising variables are coupled \emph{via} the reduced variables through the atomic Josephson energy
 \begin{equation}
- J_a \sum_{\langle r, r' \rangle}  \cos(\phi^a_{\bm r} - \phi^a_{\bm r'}) =  - \sum_{\langle r, r' \rangle} J^{\rm (eff)}_{{\bm r},{\bm r'}}~\sigma_{\bm r} \sigma_{\bm r'}\ .
\label{e.atom}
\end{equation}
 where $J^{\rm (eff)}_{{\bm r},{\bm r'}} = J_a \cos(\tilde\phi^a_{\bm r} - \tilde\phi^a_{\bm r'})$; hence generically the correlation function $\langle \sigma_{\bm r} \sigma_{\bm r'} \rangle$ is \emph{not} the one of an ordinary Ising model, unless the effective coupling $J^{\rm (eff)}$ can be considered as a static variable. 
 
\begin{figure}[t!]
\begin{center}
\includegraphics[width=1 \columnwidth]{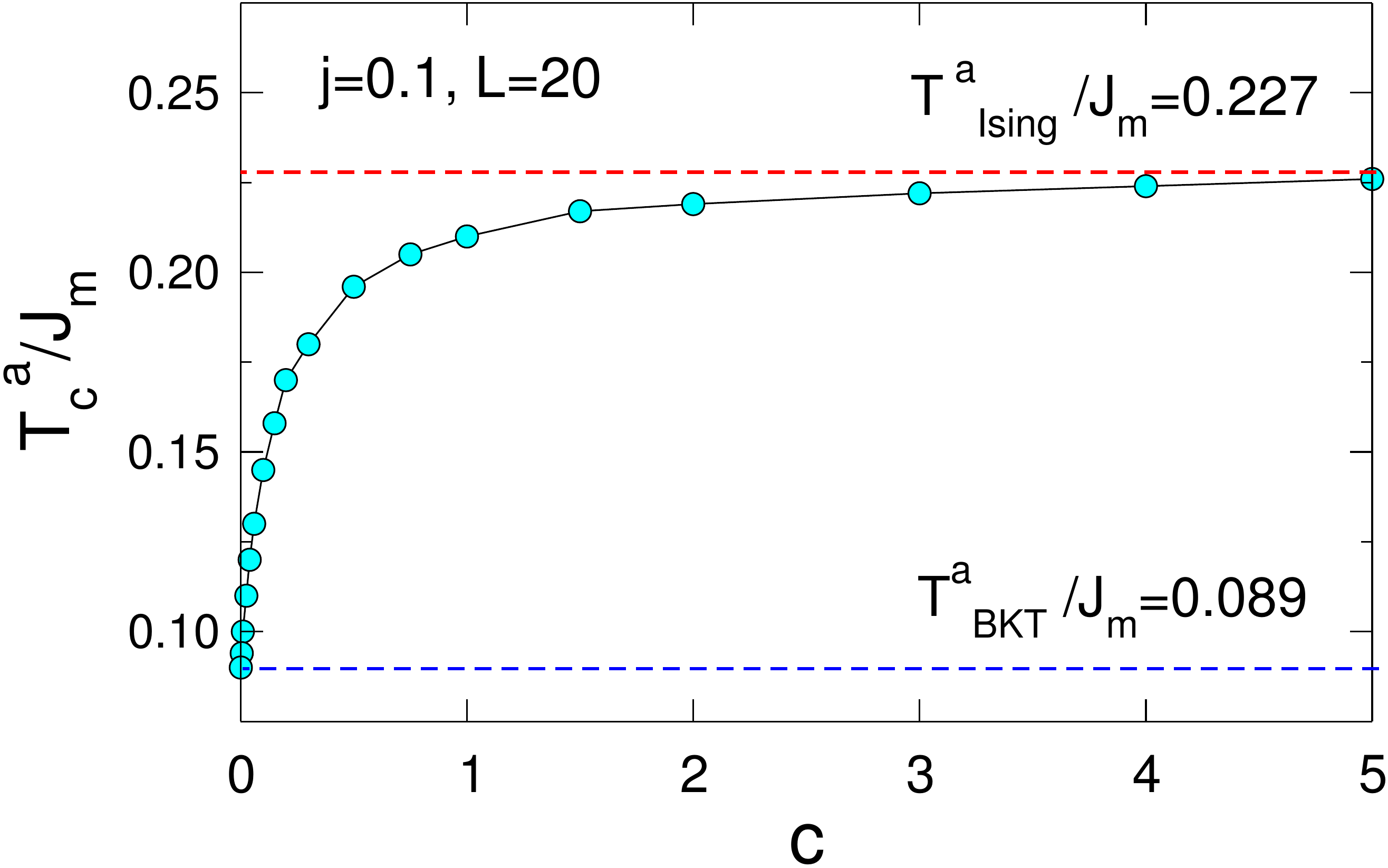}
\caption{ Atomic critical temperature $T^a_c$ for the 2$d$ Ising transition with respect to the coupling $c$ for $j$$=$$0.1$ and $L$$=$$20$.
The coupling shifts the atomic critical temperature from $T^a_{\rm BKT}/J_a\simeq0.89$ for $c \to 0$ to $T^a_{\text{Ising}}/J_a=2.27$ for $c\to \infty$.}
\label{Evolution_Tc_Ja1_Jm10_Cvarie}
\end{center}
\end{figure}

 Fig.~\ref{Evolution_Tc_Ja1_Jm10_Cvarie} shows the evolution of the atomic critical temperature $T^a_c$
 when the atom-molecule coupling $c$ increases. Under an increase of $c$ the Ising transition is seen to shift from the BKT temperature $T_{\rm BKT}^a\simeq0.89~J_a$ at $c=0$, to the critical temperature of a 2$d$ Ising model with $J_a$ spin-spin coupling, $T^a_{\text{Ising}}=(2/\ln(1+\sqrt{2}))J_a\simeq 2.27J_a$, for large $c$. A simple understanding of the large-$c$ regime can be obtained with the following consideration. When $c \gg j$, at the temperature $t \sim j (\ll c)$ where atoms quasi-condense, the phase of the molecular field is nearly homogeneous at short distances (as the molecular phase correlations decay algebraically very slowly), and the variable $2\tilde\phi^a$ is therefore locked by the very strong atom-molecule coupling to be nearly homogeneous as well. As a consequence, the only spatial dependence of the atomic phase comes essentially from the fluctuations of the Ising variable $\sigma$, $\phi_{\bm r}^a \approx \phi^m/2+\frac{1}{2}(\sigma_{\bm r}+1)\pi$. In this regime one can imagine that 
 $J^{\rm (eff)}_{{\bm r},{\bm r'}} \approx J_a$
 implying that the atomic Josephson energy, Eq.~\eqref{e.atom}, reproduces the Hamiltonian of the two-dimensional Ising model on a square lattice, whose critical temperature $T^a_{\text{Ising}}$ is attained asymptotically when $c\to\infty$. 
 
\begin{figure}[t!]
\begin{center}
\includegraphics[width=1 \columnwidth]{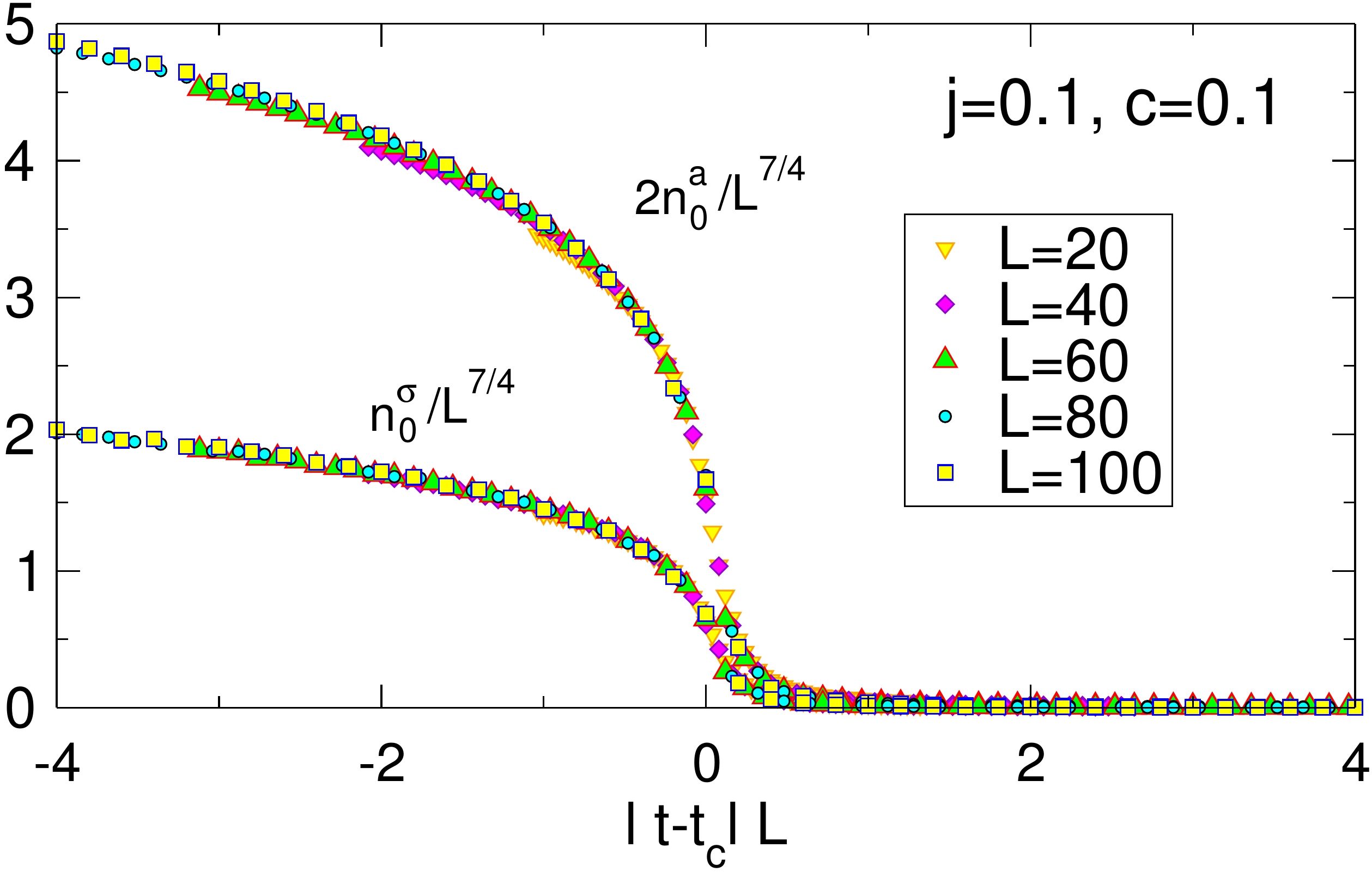}
\caption{For $j$$=$$0.1$ and $c$$=$$0.1$: collapse of the atomic quasi-condensate $n^a_0$ (multiplied by a factor of 2 for better visibility) 
and Ising structure factor $n^\sigma_0$  
using 2$d$ Ising critical exponents ($\gamma$$=$$7/4$ and $\nu$$=$$1$) with $t_c=0.153$.
}
\label{Condensates_Ja1_Jm10_L20_Cvarie}
\end{center}
\end{figure}

 As a consequence, close to (but above) the Ising critical point, the exponential decay of the Ising correlation function $\langle \sigma_{\bm r} \sigma_{{\bm r}'} \rangle \sim \exp(-|\bm r - \bm r'|/\xi_\sigma)$ induces a similar decay for $C^a_{\bm r, \bm r'}$  with same Ising correlation length $\xi_\sigma \sim |T-T_c|^{-1}$, and therefore the quasi-condensation phenomenon exhibits the unconventional Ising criticality.  
The Ising scaling of atomic correlations, which is easy to understand in the limit $c \gg j$,  remains valid also in the opposite limit $c \ll j$, as $c$ is anyway a relevant variable in the renormalization group sense, locking asymptotically the reduced atomic phase to the molecular phase. This conclusion is fully corroborated by the scaling plots Fig.~\ref{Scaling_total_na_Ja1_Jm10_CsurJm_0p01_0p1_2} and Fig.~\ref{Condensates_Ja1_Jm10_L20_Cvarie}. Fig.~\ref{Condensates_Ja1_Jm10_L20_Cvarie} in particular exhibits the same scaling plot for the atomic quasi-condensate and the Ising structure factor, showing that the criticality of the former is inherited by the latter.

\begin{figure}[t!]
\begin{center}
\includegraphics[width=1 \columnwidth]{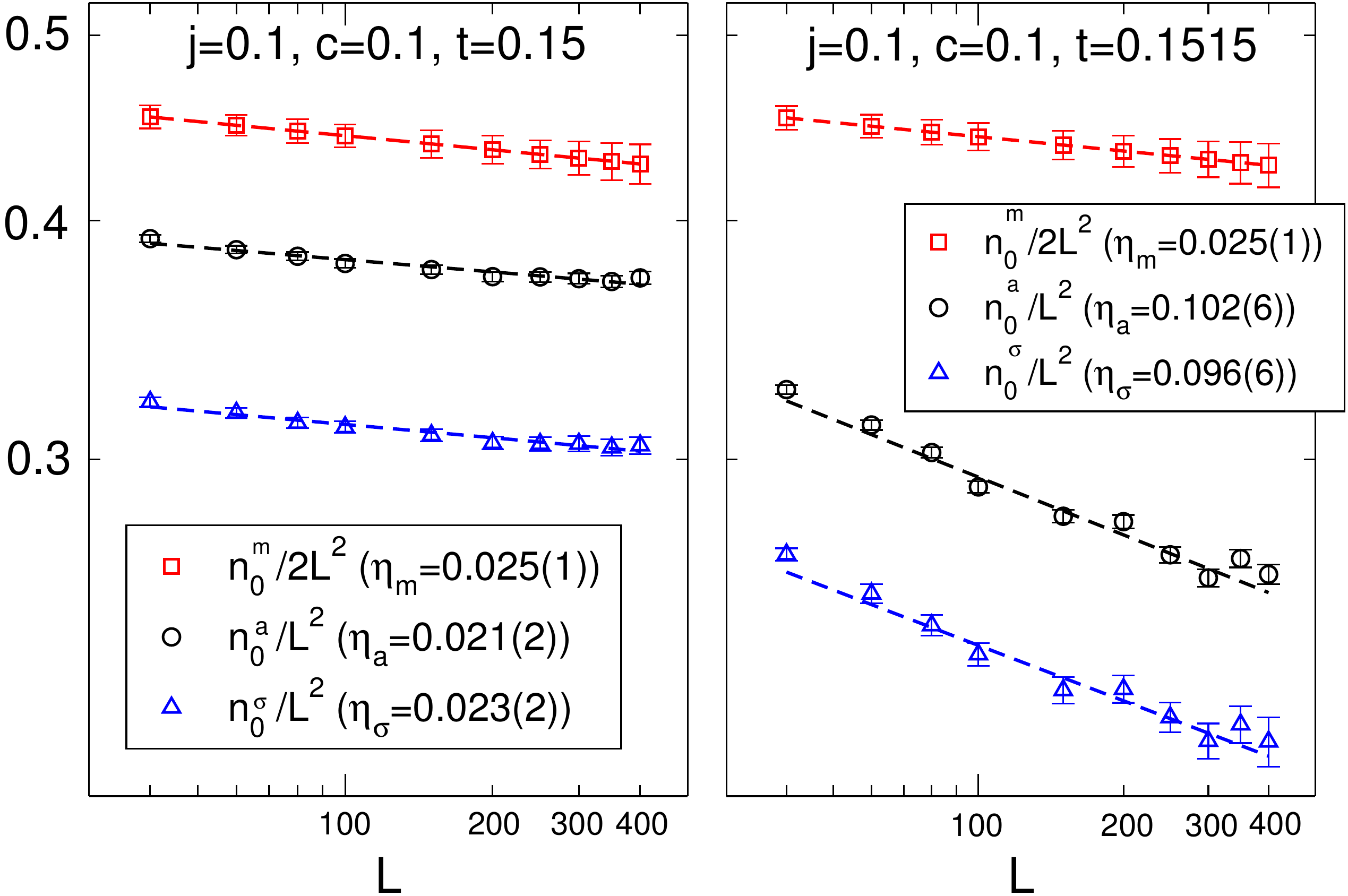}
\caption{Scaling of $n^m_0$, $n^a_0$ and $n^{\sigma}_0$ with the system size in the ${\rm SF_{am}}$ phase very close to the critical Ising temperature $t_c\simeq 0.153$, 
showing that all three vanish in the termodynamic limit, $n^\alpha_0/L^2 \propto L^{-\eta_\alpha}$ ($\alpha=a,m,\sigma$).
Note that the exponents for the Ising and atomic susceptibilities $\eta_{\sigma}$ and $\eta_a$ are already much smaller than their value at $t_c$, $\eta_{\sigma}(t_c)=\eta_a(t_c)=1/4$.}
\label{fig_scaling}
\end{center}
\end{figure}

  On the other hand, below the Ising transition the local Ising variable $\sigma$ \emph{cannot} order, as already argued in Sec.~\ref{sec_z2}, and hence it will display an algebraic decay of the kind $\langle \sigma_{\bm r} \sigma_{{\bm r}'} \rangle \propto |{\bm r}-{\bm r}'|^{-\eta_\sigma}$. The decomposition in Eq.~\eqref{e.corr_decomposition} implies then that a similar algebraic decay should be exhibited by the atomic correlations. Algebraic correlations imply quasi-ordering and quasi-condensation of Ising and atomic degrees of freedom respectively, namely  $n^a_0\propto L^{2-\eta_a}$ and $n^\sigma_0\propto L^{2-\eta_\sigma}$. These scaling forms are fully consistent with our data below the critical point, as shown in Fig.~\ref{fig_scaling}; moreover we find that the identity $\eta_\sigma = \eta_a$ is verified within error bars. Notice nonetheless that the exponent $\eta_\sigma$  decays extremely fast as soon as one leaves the critical point ($\eta_\sigma(T_c)$$=$$1/4$, while $\eta_\sigma(0.99 \,T_c)\simeq 0.1$); this is to be contrasted with the much slower thermal suppression of the $\eta$ exponent in BKT theory, see \emph{e.g.} \cite{etaXY}. This entails in turn an extremely slow spatial decay of the Ising correlations, which can be easily confused by a proper ordering of the Ising degrees of freedom, as was reported in Ref.~\onlinecite{Ejima_2011}. In Fig.~\ref{fig_scaling} we also show the scaling of the molecular quasi-condensate, $n^m_0/L^2 \propto L^{-\eta_m}$: the relationship $\eta_a = \eta_m/4$, which is claimed by Ref.~\onlinecite{Ejima_2011} assuming long-range Ising order and discards the compact nature of the angular variables, is disproved by our data.\\

  \subsection{Molecular BKT transition}

\begin{figure}
\begin{center}
\includegraphics[width=1 \columnwidth]{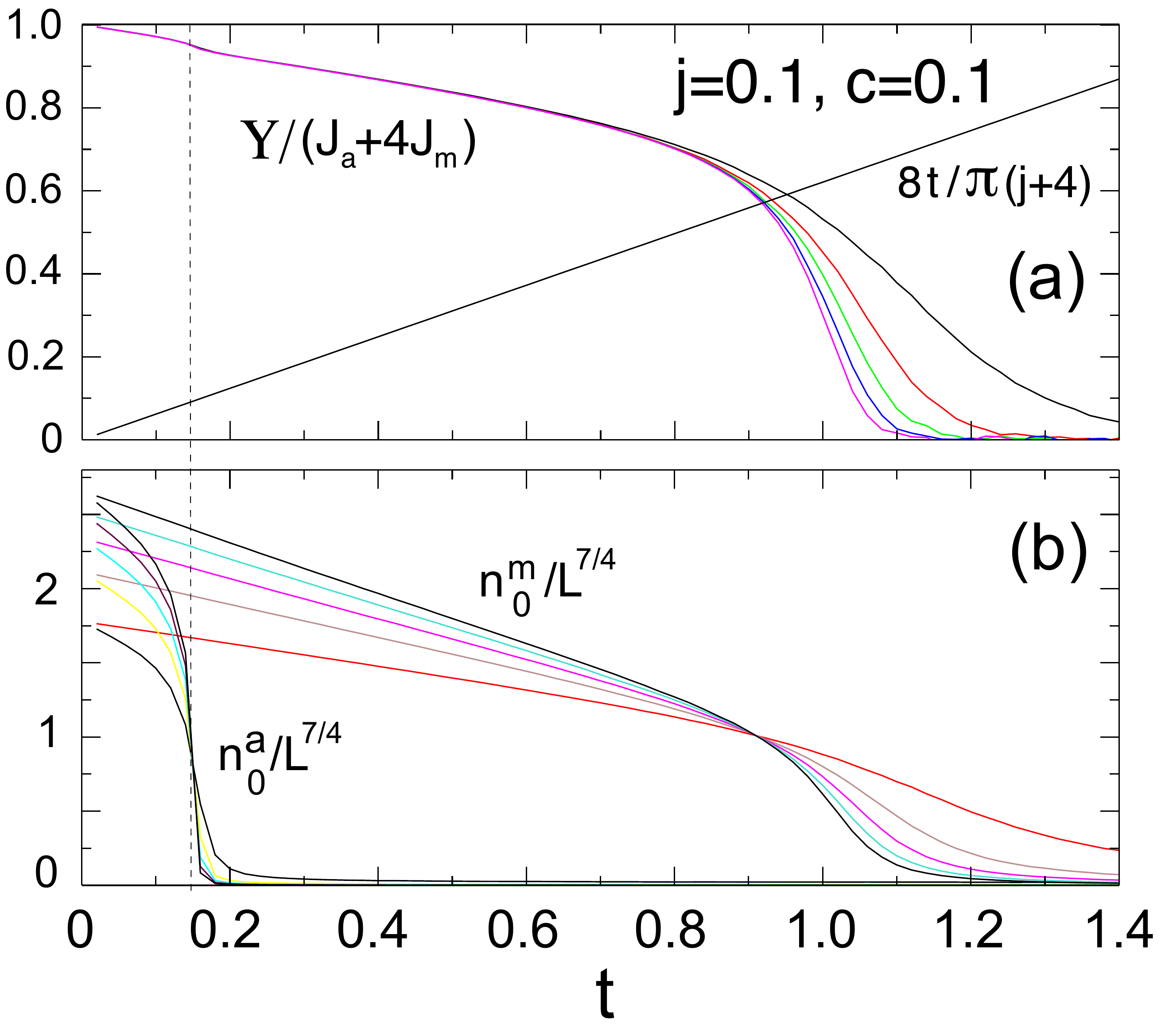}
\caption{Molecular regime $j$$=$$0.1$, and weak coupling $c$$=$$0.1$: (a)  Scaling of the spin stiffness $\Upsilon$, and (b) of the 
atomic $n^a_0$ and molecular $n^m_0$  quasi-condensates for $L$$=$$10, 20, 30, 40, 50$.
}
\label{Stiffness_Condensates_Cv_Vortex_Ja1_Jm10}
\end{center}
\end{figure}

  We now turn to the high-$T$ transition, marking the onset of molecular coherence. Our data strongly indicate that the transition is a BKT one, and we estimate its critical temperature by searching for the scaling of the molecular quasi-condensate as  $L^{2-\eta(T_{\rm BKT}^m)} = L^{7/4}$.  Fig.~\ref{Cv_scaling_condensates_2DXY_Wolff_Ja1_Jm10} (b) clearly shows that the critical BKT scaling is achieved at a temperature $T_{\rm BKT}^m\simeq 0.89 J_m$, and that its position (as well as the whole temperature dependence of the molecular quasi-condensate) is very weakly dependent on the coupling. Further inspection into the BKT transition can be achieved with the superfluid density, which is generally expected to exhibit a universal jump 
  $\Upsilon/T_{\rm BKT}=\frac2\pi$ \cite{Nelson_Kosterlitz_1977} -- and this is obviously the jump observed at the transition for $c=0$. Nonetheless, as soon as $c\neq 0$ the definition of the superfluid density changes to account for the reduced symmetry of the model, see Eq.~\eqref{spin_stiffness}, and in such a definition the phase twist for the molecular field is twice the one for the atoms. This implies that the universal jump is quadrupled (as the stiffness is a second derivative with respect to the twist), and hence unsurprisingly the BKT transition is observed to be compatible with the jump $\Upsilon/T_{\rm BKT}^m=\frac8\pi$, see Fig.~\ref{Stiffness_Condensates_Cv_Vortex_Ja1_Jm10}. 
  
\begin{figure}[t!]
\begin{center}
\includegraphics[width=0.8 \columnwidth]{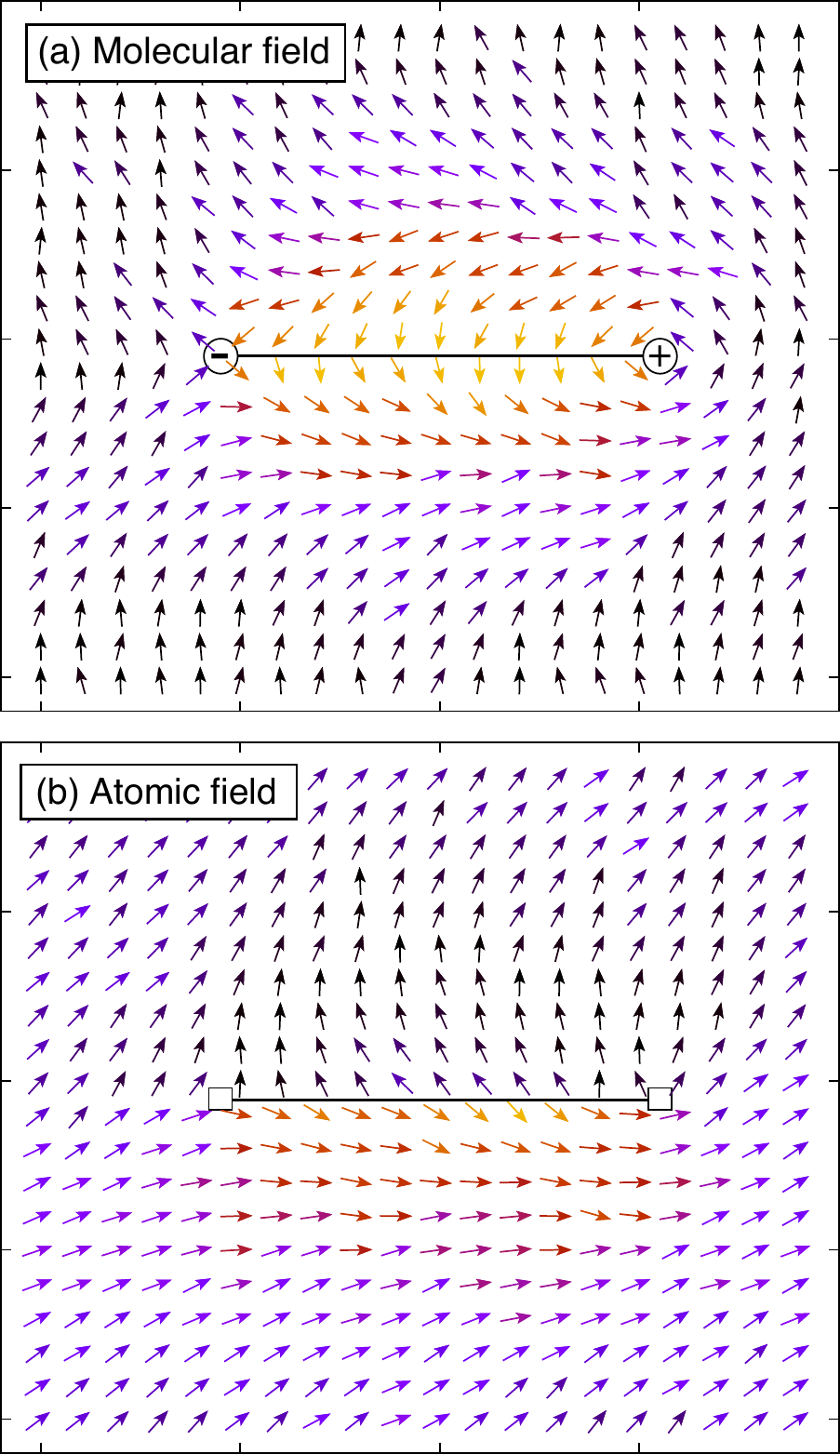}
\caption{(a) Illustration of a vortex-antivortex pair in the molecular field and (b) illustration of the corresponding half vortex-half antivortex pair in the atomic field due to the strong coupling term. The string between the two half vortices corresponds to a domain wall, whose energy grows linearly with its length.}
\label{f.halfvortex}
\end{center}
\end{figure}

 In the limit $c\to \infty$ the coupled XY-model we investigate is expected to reproduce the physics of  the $\varphi/2\varphi$ model of Eq.~\eqref{e.delta2delta}, whose superfluid stiffness has been investigated in details in Ref.~\onlinecite{HubscherWessel_2013}. In particular the atomic regime of our model corresponds to the regime of the $\varphi/2\varphi$ model dominated by the $\cos(2\varphi)$ coupling, in which two transitions (a higher-$T$ BKT transition and a lower-$T$ Ising transition) are also observed \cite{HubscherWessel_2013}. In this limit a molecular vortex translates into a vortex for the $2\varphi$ phase (or a vortex in the field of atom pairs), which is in turn a \emph{half-vortex} for the atomic field. This allows to interpret the spin stiffness jump as due to the unbinding of half-vortex/half-antivortex excitations, pictured in Fig.~\ref{f.halfvortex} (b), with an associated jump in $\Upsilon/T_{\rm BKT}^m = 2/(q^2\pi)$ with $q=1/2$. In fact the ``quadrupled" sensitivity of the spin stiffness to half-vortex excitations (or to vortices in the $2\varphi$ variable) is built into the expression of the superfluid stiffness for the  $\varphi/2\varphi$ model (Eq. 4 of Ref.~\onlinecite{HubscherWessel_2013}), as much as a similar sensitivity to molecular vortices is built in the expression of the superfluid stiffness for our double XY model, Eq.~\eqref{spin_stiffness}.  
  
\begin{figure}[t!]
\begin{center}
\includegraphics[width=1 \columnwidth]{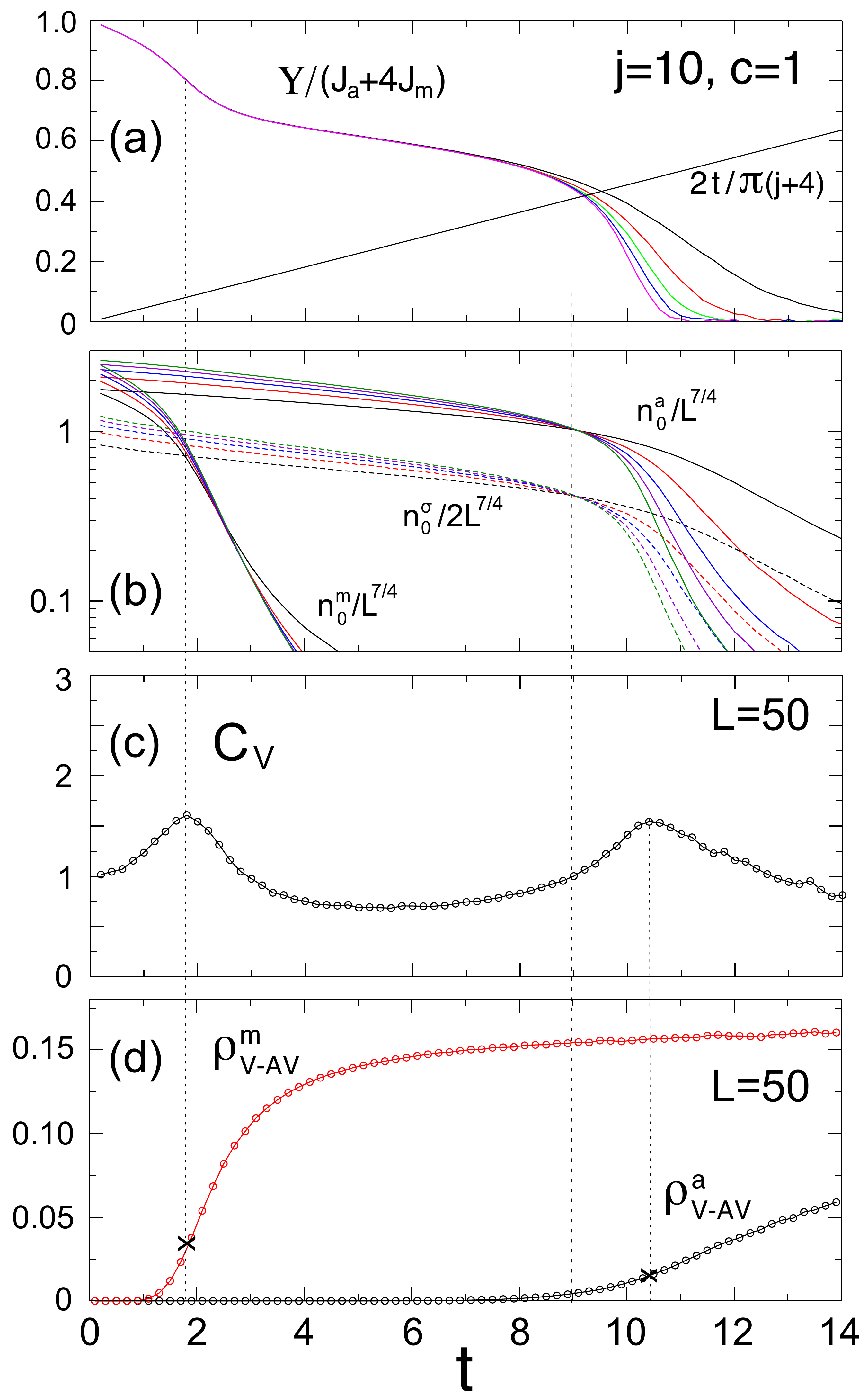}
\caption{Data for the atomic regime $j$$=$$10$, and for $c$$=$$1$: (a) Scaling of the spin stiffness $\Upsilon$, and (b) of the atomic $n^a_0$ 
and molecular $n^m_0$  quasi-condensates, and Ising structure factor  $n^\sigma_0$  in semi-log scale for $L$$=$$10, 20, 30, 40, 50$. 
For $L$$=$$50$, (c) specific heat $C_V$, and (d) atomic $\rho_{V-AV}^{a}$ and molecular $\rho_{V-AV}^{m}$  vortex densities, Eq.~\eqref{totalvorticity}.}
\label{Stiffness_Condensates_Cv_Vortex_Ja10_Jm1_C1}
\end{center}
\end{figure}

\section{Atomic regime: $j\gg 1$ }
\label{sec_JaggJm}

In the limit $j\gg 1$, we expect the build-up of correlations in the atomic field to drive any transition process when cooling down the system from high temperatures.  As we will show below, in terms of atomic correlations and superfluid response the transition is well described by a standard BKT transition at a critical temperature $T_{\rm BKT}^a$. At this transition both the atomic field, as well as the field of atom pairs, acquire algebraically decaying correlations and quasi-condense -- in particular, as discussed in the Appendix \ref{a.2phi}, the critical scaling of the atom-pair quasi-condensate is $n^{aa}_0 \sim L$.

The fate of the molecules under this circumstance has to be analyzed with care. As already discussed in Sec.~\ref{sec_symmetries}, at the mean-field level the ordering of atoms entails immediately the ordering of molecules; yet, if the atoms only exhibit quasi-LRO, the molecules see a fluctuating atom-pair field which averages to zero, and hence could be naively thought to decouple from atoms, as it would happen for $c=0$. Nonetheless this conclusion would imply that the molecules have short-range (exponentially decaying) correlations, albeit being coupled to the atom-pair field which exhibits algebraic correlations. In the limit $c \gg j$ and at temperatures $t \sim j \ll c$, this is not possible, and hence the onset of quasi-LRO of atom pairs necessarily implies a similar phenomenon for molecules, whose correlations will indeed mimic the ones of atom pairs. Hence, is there a coupling-decoupling transition between $c = 0$ and $c \to \infty$  at some non-trivial value of $c$? The answer is no, and in fact this transition occurs strictly at $c = 0$, as we will discuss below.

\subsection{Atomic BKT transition vs. molecular crossover}

Fig.~\ref{Stiffness_Condensates_Cv_Vortex_Ja10_Jm1_C1} shows that, for  $j=10$ and $c=1$, there is a transition associated with the onset of quasi-LRO of the atoms at a temperature $T_{\rm BKT}^a\simeq 0.89 J_a$, namely at a temperature very close to that of the atoms decoupled from the molecules. Indeed this transition is well described by the standard BKT scenario: at the temperature where the atomic quasi-condensate scales as $L^{7/4}$ (Fig.~\ref{Stiffness_Condensates_Cv_Vortex_Ja10_Jm1_C1}(b)), 
the superfluid stiffness is compatible with a jump of $2T_{\rm BKT}^a/\pi$ (Fig.~\ref{Stiffness_Condensates_Cv_Vortex_Ja10_Jm1_C1}(a)), associated with the binding of vortex-antivortex pairs, and a rounded peak in the specific heat appears above the transition (Fig.~\ref{Stiffness_Condensates_Cv_Vortex_Ja10_Jm1_C1}(c)). For the atomic correlation function in Eq.~\eqref{e.corr_decomposition} to exhibit algebraic decay, it is necessary that the Ising variables $\sigma_{\bm r}$ acquire at least algebraic correlations: this implies that the atomic BKT transition is associated with the concomitant development of critical correlations, as witnessed by the critical scaling of the Ising structure factor in Fig.~\ref{Stiffness_Condensates_Cv_Vortex_Ja10_Jm1_C1}(b). \\

Nonetheless, unlike in the molecular regime, the transition for the Ising $\sigma$ variables ``hidden" in the atomic BKT transition is \emph{not} of the Ising type, because the effective coupling among the Ising variables
\begin{equation}
-J\cos(\phi_{\bm r}^a - \phi_{\bm r'}^a ) = -J \cos(\tilde\phi_{\bm r}^a - \tilde\phi_{\bm r'}^a) \sigma_{\bm r} \sigma_{\bm r'} =  -J_{\rm eff}    \sigma_{\bm r} \sigma_{\bm r'}
\end{equation}
 fluctuates strongly due to the fluctuations of the reduced phase variable $\tilde\phi^a$, which, in the regime under consideration, is not locked to a weakly fluctuating molecular variable. In this case, the transition is completely controlled by the fluctuations of the $\tilde\phi^a$ variable, and the Ising variables develop the same exponentially diverging correlation length as for the angular variable, whence the BKT character of the correlation function in Eq.~\eqref{e.corr_decomposition}.
 
 On the other hand the molecular coherence is extremely weak at the atomic BKT transition, and the molecular quasi-condensate is found to be (marginally) compatible with a $L^{7/4}$ only at a much lower temperature -- therefore an ordinary BKT transition for molecules is easily ruled out in correspondence with the BKT transition for atoms. The build-up of coherence in the molecular field at a lower temperature, seen in Fig.~\ref{Stiffness_Condensates_Cv_Vortex_Ja10_Jm1_C1}(b), could be naively associated with a BKT transition decoupled from the atomic one -- also in light of a second peak observed in the specific heat at low temperature. Moreover the latter peak is associated with an inflection point of the total  molecular V-AV density $\rho_{V-AV}^{m}$, which could suggest a V-AV deconfinement phenomenon, see Fig.~\ref{Stiffness_Condensates_Cv_Vortex_Ja10_Jm1_C1}(d). 
In fact all these observations turn out to be rather misleading, and a deconfinement transition is easily ruled out by analyzing the temperature and size dependence of the superfluid stiffness. As shown in Fig.~\ref{Stiffness_Condensates_Cv_Vortex_Ja10_Jm1_C1}(a), the buildup of molecular coherence and the peak in the specific heat are associated with a rounded shoulder in the stiffness, which does \emph{not} show any sign of size dependence, hence ruling out the existence of a jump.

\begin{figure}[t!]
\begin{center}
\includegraphics[width=1 \columnwidth]{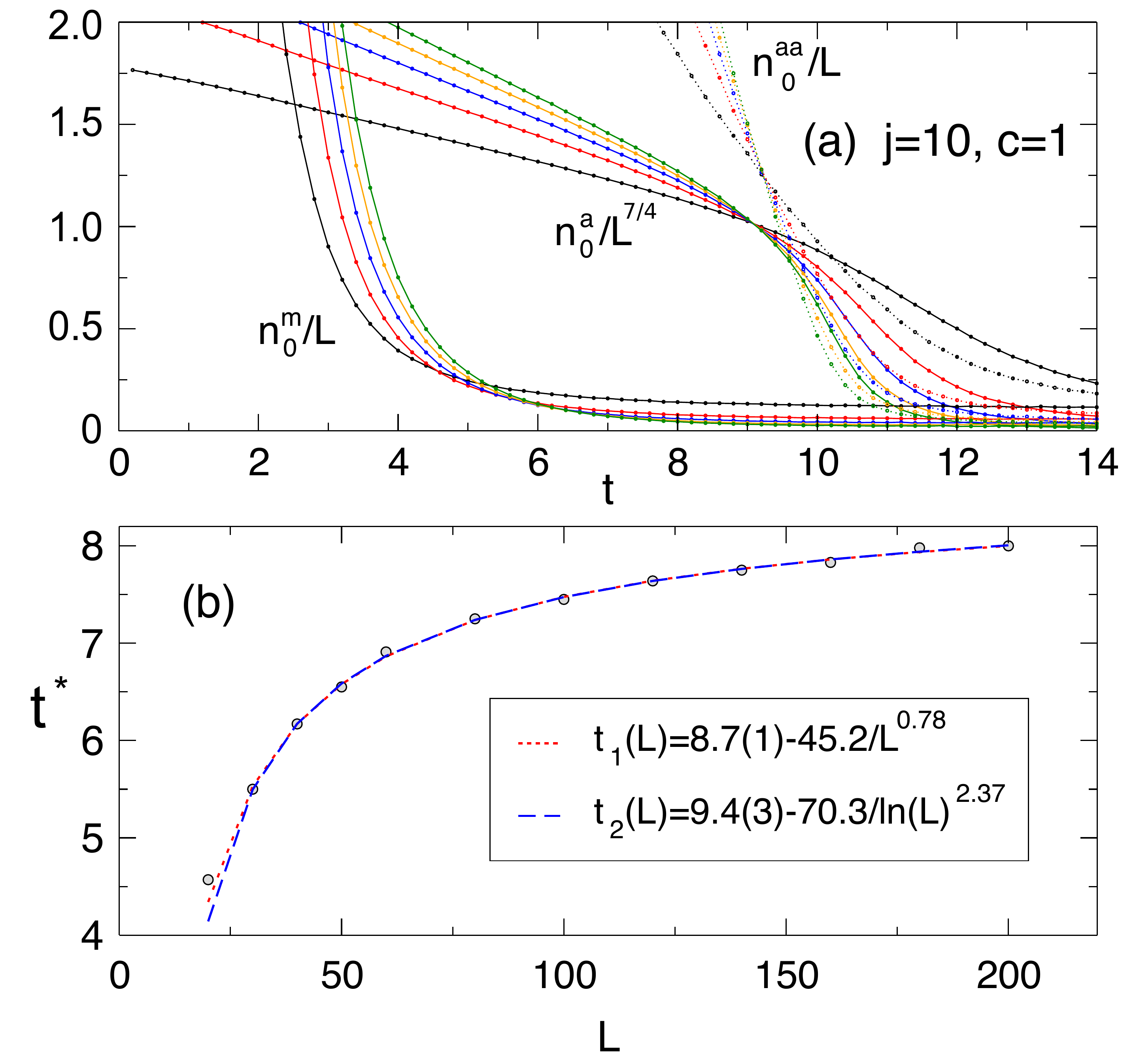}
\caption{(a) Scaling of atom pairs $n^{aa}_0$, 
atomic $n^a_0$ and molecular $n^m_0$  quasi-condensates for $L=10, 20, 30, 40, 50$;
(b) The temperature $t^*$ at which $n^m_0$ exhibits a linear scaling in $L$ slowly drifts towards the atomic BKT transition at $t_{BKT}^a$$\simeq$$8.9$ 
as $L$ increases. $t_1(L)$ and $t_2(L)$  are two possible fits for the evolution of $t^*(L)$, which are compatible with $t_{BKT}^a$ in the thermodynamic limit.
}  
\label{f.scaling}
\end{center}
\end{figure}

\begin{figure*}[t!]
\begin{center}
\includegraphics[width=1 \columnwidth]{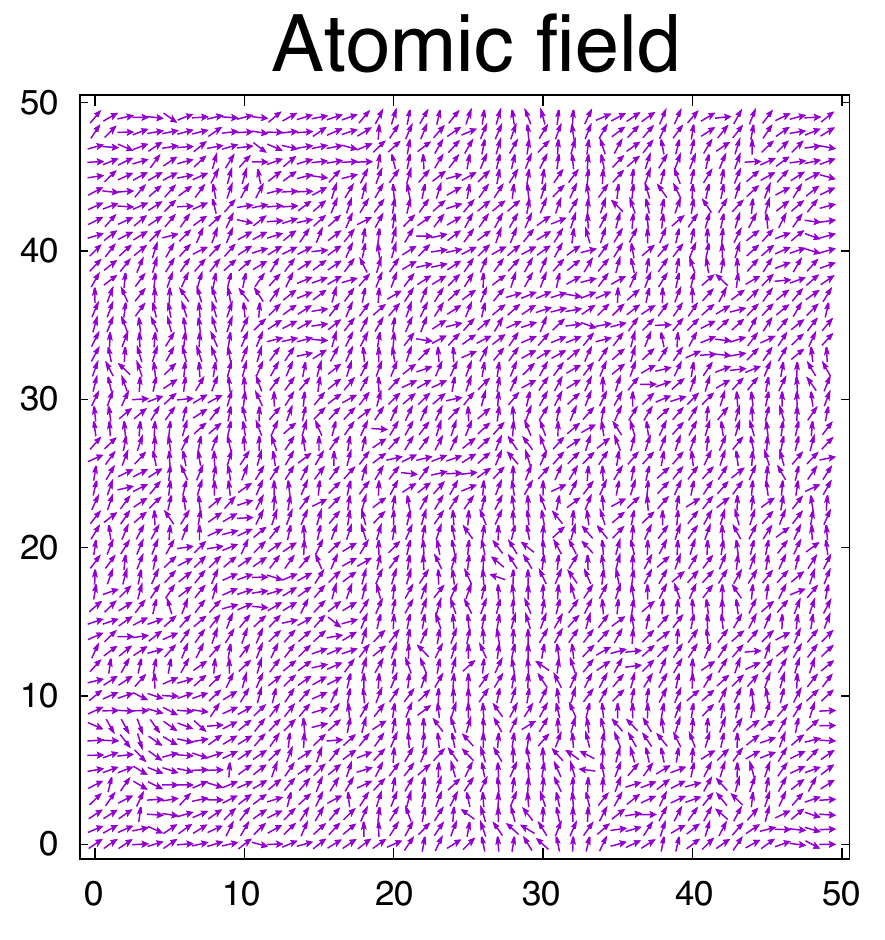}
\includegraphics[width=1 \columnwidth]{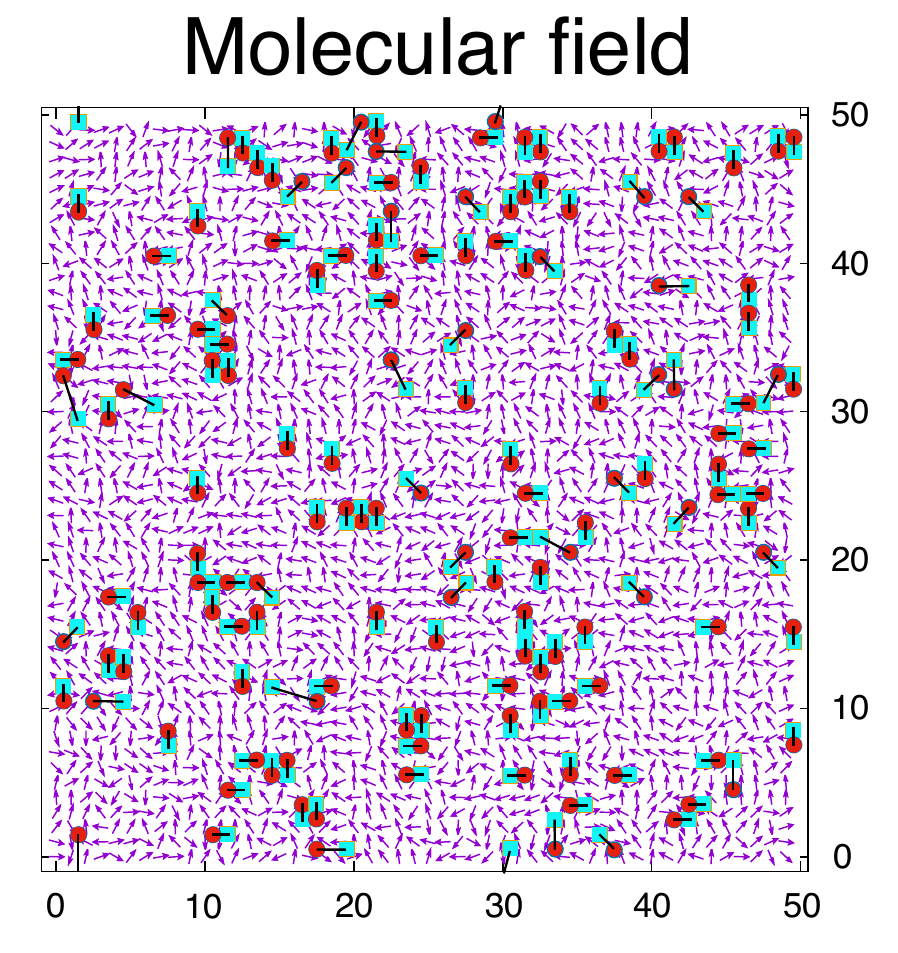}
\caption{Snapshot of the atomic and molecular fields in the SF$_{am}$ phase of the atomic regime ($j$$=$$10$, $ c$$=$$5$) for a temperature $t$$=$$4$ 
which is above the molecular crossover (see text). 
There are no vortices in the atomic field, while there are many bound V-AV pairs in the molecular field.
The local vorticity $\rho_{V,\Box}^{m}$ coded in color as red/blue for a vortex/antivortex; the black lines identify the bound V-AV pairs. 
}
\label{f.snapshot}
\end{center}
\end{figure*}  
  
\subsection{Nature of the molecular quasi-condensation and molecular crossover}

 How to reconcile the low-temperature entropy release associated with the increase in the molecular vorticity, and the absence of any transition in the molecular field? 
 A first fundamental observation comes from the fact that the angular variable $\Omega = 2\phi^a - \phi^m$ is coupled to the static ``magnetic field" $C$ in the Hamiltonian Eq.~\eqref{XY_hamiltonian}, and as such it is always ordered at any temperature, namely $\langle \cos(\Omega_{\bm r}) \rangle \neq 0$, see also Appendix \ref{a.naam}. This immediately implies that the onset of algebraic correlations in the variable $2\phi^a$ necessarily entails the exact same form of correlations of the variable $\phi^m$, as the long-wavelength behavior of the two variables is necessarily the same. This is true regardless of the magnitude of $c$: as already observed above, the atom-molecule coupling is a relevant variable in the renormalization-group sense, because it reduces the symmetries of the system, and as such it is always diverging along the renormalization-group flow. This implies that, along the renormalization-group flow, the molecular phase correlates perfectly with the variable $2\phi^a$, mimicking its algebraic correlations not only at the atomic BKT transition, but throughout the BKT phase at low temperature. Hence the molecular field is expected to quasi-condense at the atomic BKT transition in the same way as atom pairs condense, namely with $n^m_0 \sim L$. Fig.~\ref{f.scaling} suggests that this is indeed the case when $L \to \infty$: the temperature at which $n^m_0$ exhibits a linear scaling in $L$ slowly drifts towards the atomic BKT transition as $L$ increases, consistent with the increase of the coupling $c$ along the renormalization-group flow.
Indeed, an extrapolation of the temperature realizing a linear scaling of $n^m_0$ to the limit $L\to\infty$ indicates that the temperature realizing an asymptotic linear scaling is consistent with the atomic BKT transition, see Fig.~\ref{f.scaling}(b).  However, the length-scale beyond which we expect this scaling to appear can be very large for small $c$, as it is related to the confinement length of the molecular vortices (which diverges as $c\to0$), see discussion below. This explains the need for the infinite-size extrapolation in Fig.~\ref{f.scaling}(b). However, for $c\gg1$, one easily finds the expected scaling $n^m_0 \sim L$ at $t_{\rm BKT}^a$ without the need for extrapolations, since $n^m_0$ mimics the behavior of the atomic pairs exactly (not shown).
    
  The above picture implies that the molecular V-AV excitations bind already at the atomic BKT transition occurring at a temperature $t\sim j$, despite the fact that they seem to persist in the thermodynamics of the system down to a much lower temperature scale $t\sim 1$. Hence within the SF$_{am}$ phase in the atomic regime, the molecular field has the peculiar nature of displaying a high density of weakly bound V-AV dipoles. This is clearly shown in Fig.~\ref{f.snapshot}, presenting a snapshot of a MC simulation in the regime of bound V-AV molecular dipoles.

A more microscopic way of understanding the binding of molecular vortices below the atomic BKT transition is to evaluate the energetic cost of their unbinding. In the case $c=0$, the standard argument to evaluate the free-energy cost of a free molecular vortex leads to the estimate
 \begin{equation}
\Delta F^m_V(C=0)\simeq (\pi J_m-2T)\ln L\ ,
\end{equation}
suggesting a transition at a temperature $T \sim \pi J_m/2$. Yet this argument has to be revised when $c>0$, because the appearance of an unbound vortex and antivortex, separated by a distance $l$, leads to a deformation of the phase configuration in the molecular field which extends typically over a region of size $l \times w$ (where $w$ is some appropriate width). For $c\ll j$, and if $T$ is below the atomic BKT transition, namely if the atomic field does not contain free vortices, the energy cost for the unbinding of a molecular V-AV pair has to be corrected to give
 \begin{equation}
\Delta F^m_{V-AV}(C)\simeq 2(\pi J_m-2T)\ln L + C w l~.
\end{equation}
Hence the atomic field provides a confinement force which will have the tendency to bind the vortices over a distance $\bar{l}$ such that  $\bar{l} \sim T/ (Cw)$. 
 This additional linear binding force is always confining, even at temperatures at which the intrinsic logarithmic interaction among molecular vortices ceases to be confining (namely at temperatures exceeding the BKT transitions of molecules decoupled from the atoms).

Alternatively, if $c\gg j$, the atomic field would be forced to minimize the coupling energy with the molecular one by developing a half V-AV pair in correspondence with the V-AV excitations of the molecules, with an additional energy cost of 
 \begin{equation}
\Delta F ^a_{1/2}\simeq \frac{\pi J_a}{2} \ln L + (2J_a - T \gamma) l
\end{equation}
where the first term comes from the long-range deformation in the atomic phase configuration, and the second one from the string connecting the two half vortex excitations. If the atomic half V-AV pair is to mimic closely the configuration of the molecular vortices, there is no further entropic gain coming from the position of the half-vortex cores, but only from the geometry of the string, see Fig.~\ref{f.halfvortex}, which we evaluate as $\gamma l$ for a string of length $l$, where $\gamma$ is some constant of order ${\cal O}(1)$. Hence, unless $T  \gtrsim J_a$, the string tension tends to confine the half vortices, and moreover the logarithmic energy cost of the half vortices would have to shift the unbinding transition of the molecular vortices by a term $\sim J_a$.  Hence in both cases we see that the atom-molecule coupling has the effect of binding the molecular V-AV pairs for temperature $T \lesssim J_a$.

\begin{figure}[ht!]
\begin{center}
\includegraphics[width=1 \columnwidth]{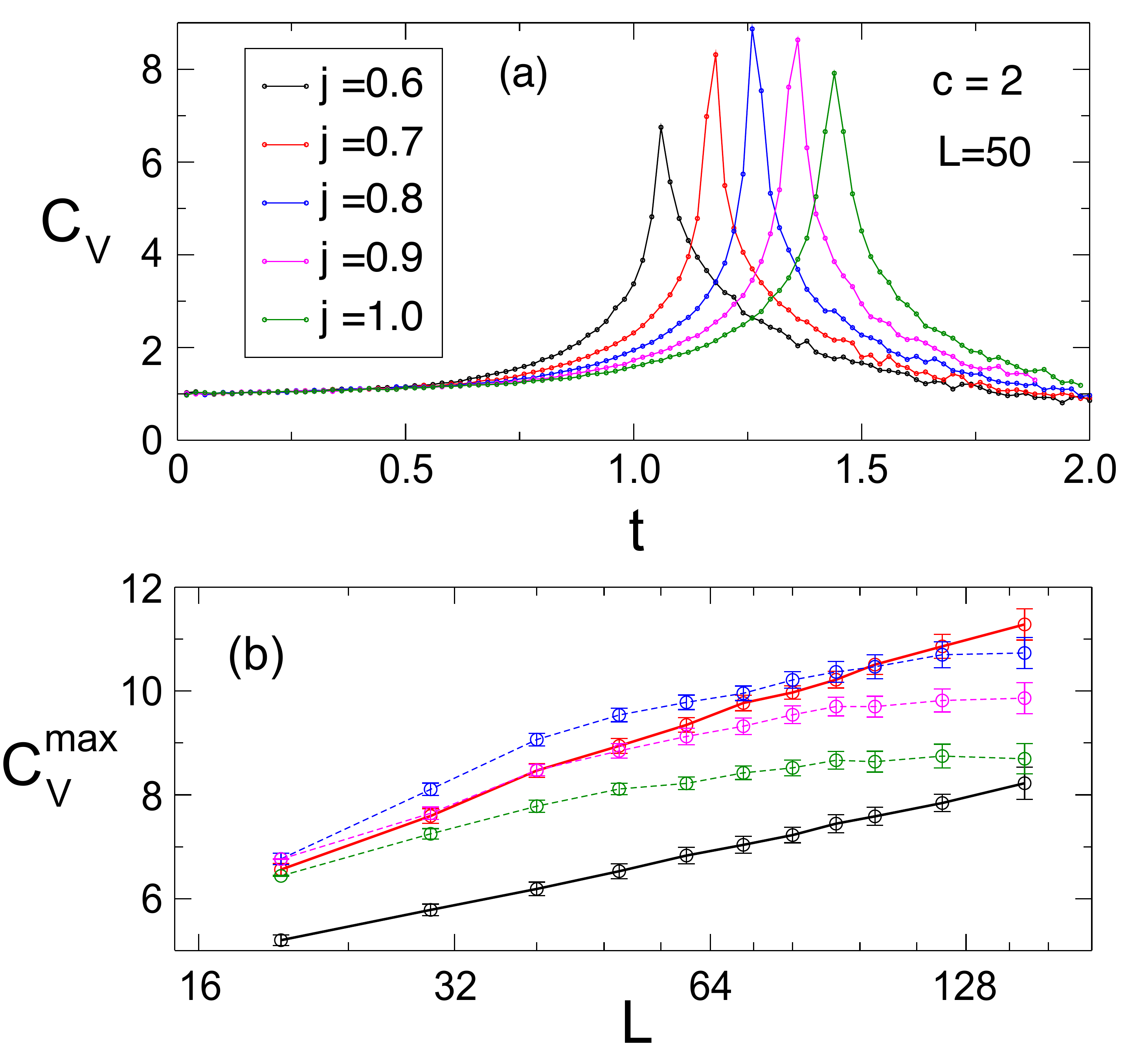}
\caption{ (a) Specific heat $C_V$ for $c$$=$$2$ and $j$$=$$0.6, 0.7, 0.8, 0.9, 1$ (b) scaling of its maximum $C_V^{max}$ as a function of $L$ (semi-log scale) for the same values of the parameters. 
$C_V$ seems to diverge only for $j$$=$$0.6, 0.7$.
}
\label{Cv_2DXY_Wolff_Ja0p6to1_Jm1_L50_finite_size_scaling_Cv}
\end{center}
\end{figure}

\section{Resonant regime: $j\sim 1$}    
 \label{sec_JasimeqJm} 

We have seen that the molecular regime $j \ll 1 $ is characterized by two successive transitions, while the atomic regime $j \gg 1$ possesses only one transition. This implies that the two transition lines in the former regime have to merge at a tricritical point for $j \sim 1$. The physics close to this tricritical point requires a careful treatment, as possible strong crossover effects are expected there.
Using a 2$d$ renormalization group analysis,  Granato, Kosterlitz and Poulter have argued the possibility of a first-order transition close to the tricritical point when there is a single transition \cite{Granato_1986,erratum_Granato}. A more recent study by Shahbazi and Ghanbari, using Monte Carlo simulations of relatively small sizes, reported a violation of the universality hypothesis, with critical exponents varying with the coupling constants in this parameter range $j\sim c \sim 1$ \cite{Shahbazi_2006}.
As we will now show, neither of these conclusions are in agreement with our results, obtained for larger system sizes than in Ref.~\onlinecite{Shahbazi_2006}.

 \begin{figure}[t!]
\begin{center}
\includegraphics[width=1 \columnwidth]{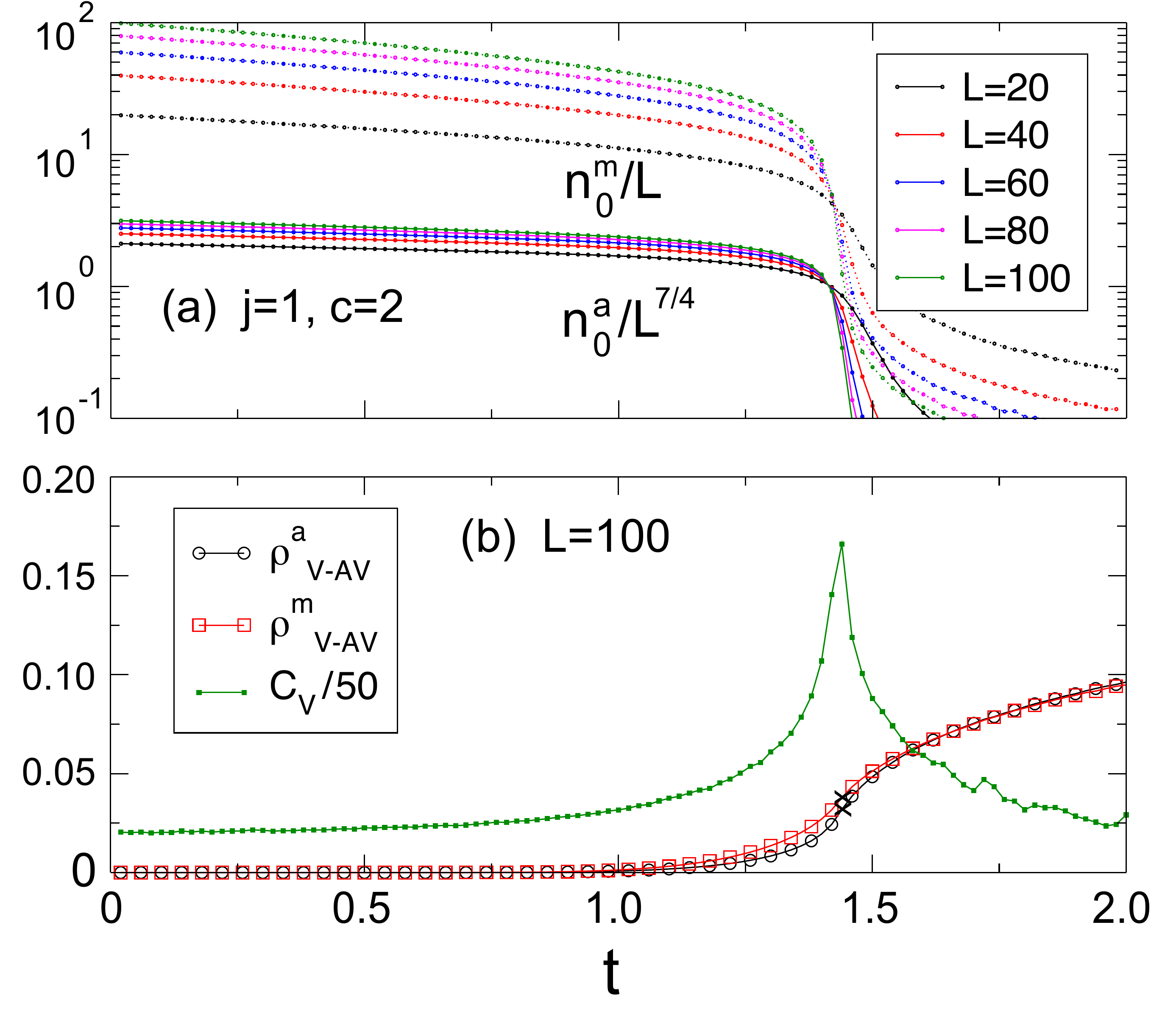}
\caption{ For $j$$=$$1$ and $c$$=$$2$: (a)  $n^a_0/L^{7/4}$ and $n^m_0/L$ in semi-log scale 
for $L$$=$$20, 40, 60, 80, 100$; (b) For $L$$=$$100$,  $C_V$ divided by 50 for better visibility and atomic and  molecular vortex densities, Eq.~\eqref{totalvorticity}.}
\label{vortex_density_scaling_na_Cv_Ja1_C2}
\end{center}
\end{figure}

We start with the case $c$$=$$2$, and varying $j\in[0.6,1]$, which is the precise range of parameters studied in Ref.~\onlinecite{Shahbazi_2006}. The specific heat is shown in Fig.~\ref{Cv_2DXY_Wolff_Ja0p6to1_Jm1_L50_finite_size_scaling_Cv} (a) for a system size $L=50$. There is a single clear maximum for each $j$ value, which is pushed to higher temperature as $j$ increases. Fig.~\ref{Cv_2DXY_Wolff_Ja0p6to1_Jm1_L50_finite_size_scaling_Cv} (b) shows the scaling of the maximum with system size: we clearly see that a logarithmic divergence of the maximum for $j \lesssim 0.7$, characteristic of the Ising atomic transition on the molecular side, leaves the place to a non-divergent maximum for $j \gtrsim 0.8$, consistent with a BKT scenario. It is important to notice that large system sizes ($L> 60$) are necessary to observe the saturation of the specific heat maximum for $j \gtrsim 0.8$. This might explain why Ref.~\onlinecite{Shahbazi_2006}, limited to sizes $L\leq 60$, was able to fit the slow (non-universal) rise of the maxima with non-universal critical exponents. Fig.~\ref{vortex_density_scaling_na_Cv_Ja1_C2} shows that, for $j=1$, the atomic quasi-condensate $n^a_0$ scales as $L^{7/4}$ and the molecular quasi-condensate as $L$ at a critical temperature very close to the position of the maximum of $C_V$, and that close to the same temperature both atomic and molecular vorticities display a net increase -- both observations corroborate a conventional BKT scenario, already preluding the atomic regime.
 
\begin{figure}
\begin{center}
\includegraphics[width=1 \columnwidth]{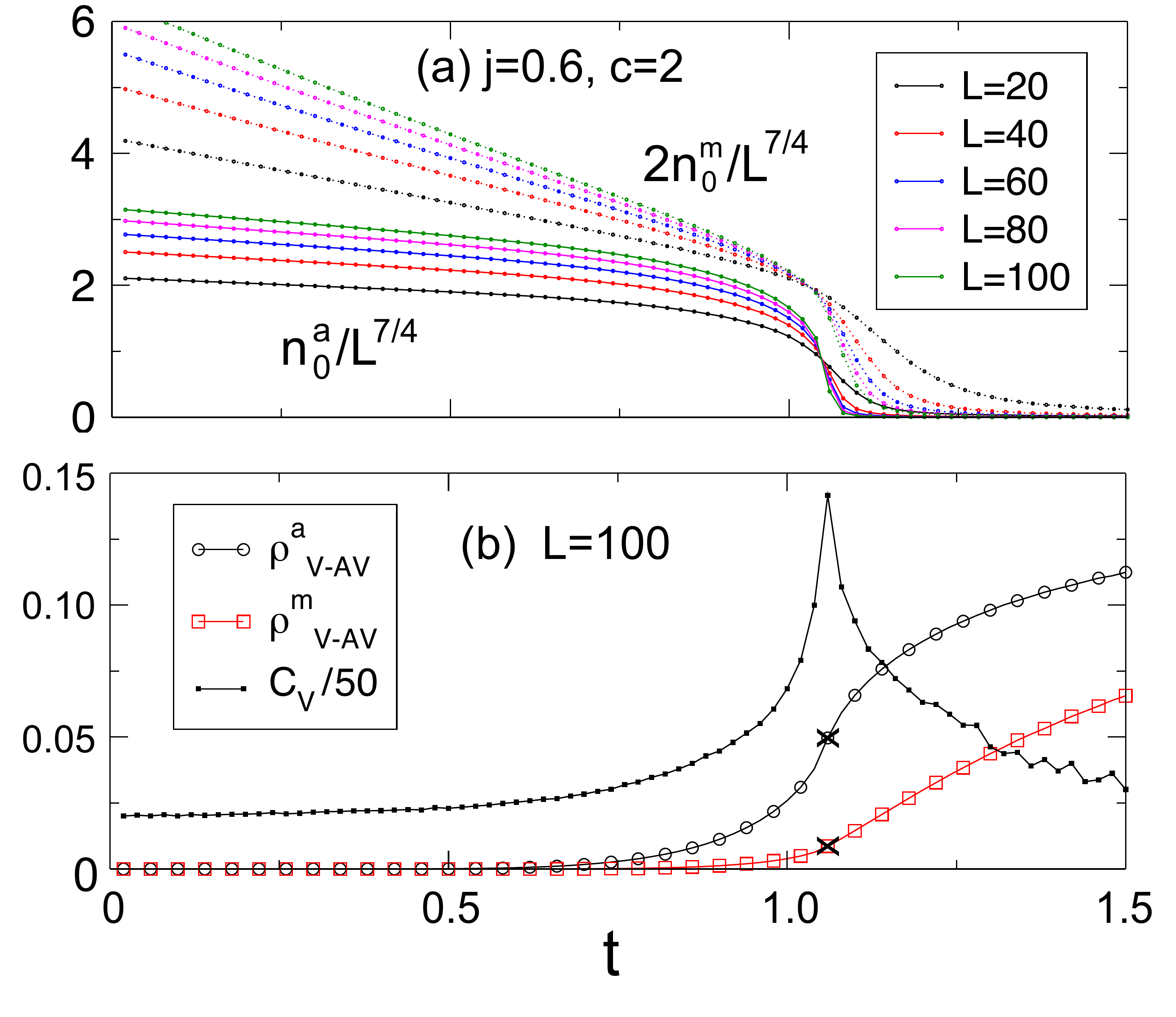}
\caption{For $j$$=$$0.6$ and $c$$=$$2$: (a)  $n^a_0/L^{7/4}$ and $n^m_0/L^{7/4}$, multiplied by  2 for better visibility,
for $L$$=$$20, 40, 60, 80, 100$; (b) For $L$$=$$100$,  $C_V$ divided by 50 for better visibility and atomic and  molecular vortex densities, Eq.~\eqref{totalvorticity}.}
\label{vortex_density_scaling_na_Cv_Ja0p6_C2}
\end{center}
\end{figure}

On the other hand, we find for $j \lesssim 0.7$ a  logarithmically diverging peak in the specific heat, see  Fig.~\ref{Cv_2DXY_Wolff_Ja0p6to1_Jm1_L50_finite_size_scaling_Cv}, as well as scalings of the atomic and molecular quasi-condensates that are consistent with an atomic Ising transition, almost coinciding with the molecular BKT transition, see Fig.~\ref{vortex_density_scaling_na_Cv_Ja0p6_C2}.  
Yet, to acquire its Ising character, the atomic transition must occur at a slightly lower temperature than the molecular transition, at which the fluctuations of the reduced phase variable ${\tilde\phi}^a$ are suppressed and the fluctuating Ising variable $\sigma$ emerges.
  This indicates that the tricritical point is somewhere between $j$$=$$0.7$ and  $j$$=$$0.8$ for $c$$=$$2$, but a detailed investigation of the position and nature of the tricritical point is beyond the scope of our present work.

\section{Conclusions}
\label{conclusion}
We unveiled the universal traits of the finite-temperature transitions in coherently coupled 2$d$ atom/molecule mixtures by numerically studying a classical XY model fully capturing the symmetries of the phase degrees of freedom of the mixture. The model parameters are related to experimentally controllable microscopic parameters of the full Hamiltonian, namely the atomic and molecular densities or effective masses (in a lattice), and the coherent coupling between the two species, achieved close to a Feshbach resonance or via photo-association. When the mixture is strongly imbalanced in favor of molecules, the onset of quasi-condensation in the molecules is decoupled from that of the atoms because of a hidden Ising degree of freedom associated with the phase of the atomic field. This results in two separate transitions for molecular quasi-condensation and for atomic quasi-condensation, the latter one having an unusual Ising nature,  and the even more unusual feature that no obvious form of long-range order can be identified below it. 
In the opposite limit of strong imbalance in favor of atoms, the two transitions merge into a single one with conventional BKT character for atoms, although the quasi-condensation of molecules is anomalous, being \emph{de facto} induced by the quasi-condensation of atomic pairs. This leads to a very weak quasi-condensate with a sizable density of weakly bound vortex-antivortex excitations, crossing over to strongly bound pairs and strong quasi-condensation only at a much lower temperature. Our conclusions are supported by extensive Monte Carlo simulations based on a newly introduced Wolff algorithm, which can deal with couplings of the general form $\cos(\phi^m_{\bm r} - p\,\phi^a_{\bm r})$ where $p$ is an integer. The relevance of this approach goes far beyond the study of many-body physics for cold atoms: as an example, the smectic-A--hexatic-B transition in liquid crystals \cite{Bruinsma_1982} is described by the above coupling with $p=3$. 

 As for cold-atom physics, all this phenomenology sets the stage for future experiments on coherently mixtures of atomic and (collisionally stable) molecules, showing the vast richness in the many-body physics which results from the quantum coherence established between two different particle numbers (\emph{two} atoms and \emph{one} molecule). The finite-temperature phase diagram bears substantial analogies with the one predicted for one-dimensional mixtures at zero temperature \cite{Ejima_2011, Bhaseenetal2012}, where an Ising transition for atomic quasi-condensation is also observed. A finite-temperature and two-dimensional realization of this physics has obvious advantages for the experimental feasibility, and moreover it bears the potential to unveil the unconventional 2$d$ topological excitations appearing in the system, namely weakly bound molecular vortex dipoles and half atomic vortices, which can be potentially imaged via the interference of independently prepared atomic clouds \cite{Hadzibabicetal2006, Chomazetal2014}.

\begin{acknowledgments}
We thank B. Delamotte, F. H\'ebert, P. Hohenberg, P. Holdsworth, G. Tarjus, and L. Tarruell for useful discussions. This work is supported by ANR (``ArtiQ" project). All calculations have been performed on the PSMN center of the ENS-Lyon. 
\end{acknowledgments}

\appendix
\section{Wolff algorithm for coupled XY models \label{app_wolff}}

We consider the general case of coupled XY models of the form
\begin{eqnarray}
 \nonumber 
\mathcal{H}_{\rm p}  =   &-& J_a \sum_{\langle {\bm r}, {\bm r'} \rangle}  \cos(\phi^a_{\bm r} - \phi^a_{\bm r'})   \\
  \nonumber 
  &-& J_m \sum_{\langle {\bm r}, {\bm r'} \rangle}  \cos(\phi^m_{\bm r} - \phi^m_{\bm r'})   \\
  &-& C   \sum_{{\bm r}}  \cos(\phi^m_{\bm r} - p\,\phi^a_{\bm r})~,   
\label{XY_p}
\end{eqnarray}
where $p$ is an integer. 
 
 The conventional Wolff algorithm \cite{Wolff_1989} for the XY model chooses a random direction $\theta$ on the unit circle, and grows a cluster by 1) flipping the component of a randomly picked spin which is orthogonal to the direction in question:
 \begin{equation}
 \phi  \to \phi' = -\phi + 2\theta~;
 \label{e.flip}
 \end{equation}
 2) calculating the energy cost $\Delta E$ of the flip for each bond involving the flipped spin(s); 3) adding neighboring spins to the cluster by flipping their orthogonal component with a probability $P = 1-{\rm min}[1,\exp(-\beta \Delta E)]$; 4) going back to step (2) until the cluster growth stops. 
 
This algorithm is not directly applicable to the model in Eq.~\eqref{XY_p} because of the $C$ term in the Hamiltonian: the flip operation in Eq.~\eqref{e.flip}, when performed on both the atomic and molecular phases, 
does not leave the coupling energy unchanged, so that the inclusion of a spin in the cluster via a $C$-term coupling does not restore the coupling energy to its initial value.   
 This problem has prevented all previous studies from exploiting the power of cluster algorithms in the study of models of the kind of Eq.~\eqref{XY_p}. But this obstacle can be easily surpassed by ``asymmetrizing" the algorithm between atoms and molecules, and flipping the spin corresponding to atoms with respect to the modified direction $\theta/p$. Indeed it is easy to see that the spin-flip prescription:
 \begin{equation}
\begin{split}
(\phi^{m})'&=-\phi^m+2\,\theta\ , \\
(\phi^{a})'&=-\phi^a+\frac2p\,\theta \ 
\end{split}
\label{eq_flip}
\end{equation} 
leaves the $C$-term unchanged once it is performed both on the atomic and molecular phase,  $\cos(\phi^m_{\bm r} - p\,\phi^a_{\bm r}) =  \cos[(\phi^m_{\bm r})' - p\,(\phi^a_{\bm r})']$.
For the algorithm to be ergodic, it is important that the angle $\theta$ is chosen within the interval $[0,p\pi]$ to allow the atomic spin to visit all possible configurations. While the spin-flip operation on molecules is insensitive to the shift $\theta \to \theta + k\pi$ with integer $k$, this is not the case for the atoms, which, under the $k\pi$ shift of the $\theta$ direction, transform as $(\phi^a)' \to (\phi^a)' +\frac {2\pi}{p}\, k$. When introducing a $p$-state discrete variable $I =  0, ..., p-1$ (such that $\sigma = 2I-1$ for $p=2$), one can write the atomic phase as $\phi^a = \tilde\phi^a + \frac{2\pi}{p}~I$, where $\tilde\phi^a$ is a reduced phase variable in $[0,2\pi/p]$. The $k\pi$ shift allows then to explore different values of $I$, namely it naturally takes into account the $\mathbb{Z}_p$ degrees of freedom of the model. 

\section{Scaling of the atom-pair quasi-condensate}
\label{a.2phi}

 In the spin-wave regime of the XY model, in which $- J\sum_{\langle {\bm r}{\bm r'} \rangle }\cos(\phi_{\bm r} - \phi_{\bm r'}) \approx \frac{J}{2} \int d^2 r ~(\nabla \phi)^2 + {\rm const}$ with unbounded $\phi_{\bm r}$ field, we have that 
\begin{eqnarray}
\langle (\psi^{\dagger}(\bm r))^2 \psi(\bm r')^2 \rangle \sim \langle e^{i2(\phi_{\bm r'}-\phi_{\bm r'})} \rangle \nonumber \\
= e^{-2 \langle (\phi_{\bm r'}-\phi_{\bm r'})^2 \rangle} \sim \langle \psi^{\dagger}(\bm r) \psi(\bm r') \rangle^4
\end{eqnarray} 
which implies that, at the BKT transition, 
\begin{equation}
\langle \cos[2(\phi_{\bm r} - \phi_{\bm r'})]\rangle \sim \frac{1}{|{\bm r} - {\bm r'}|^{4\eta(T_{\rm BKT})}} = \frac{1}{|{\bm r} - {\bm r'}|}~.
\end{equation} 
As a consequence, at the BKT transition the atom-pair quasi-condensate scales as $n^{aa}_0\sim L$.

\begin{figure}[ht!]
\begin{center}
\null\bigskip
\includegraphics[width=\columnwidth]{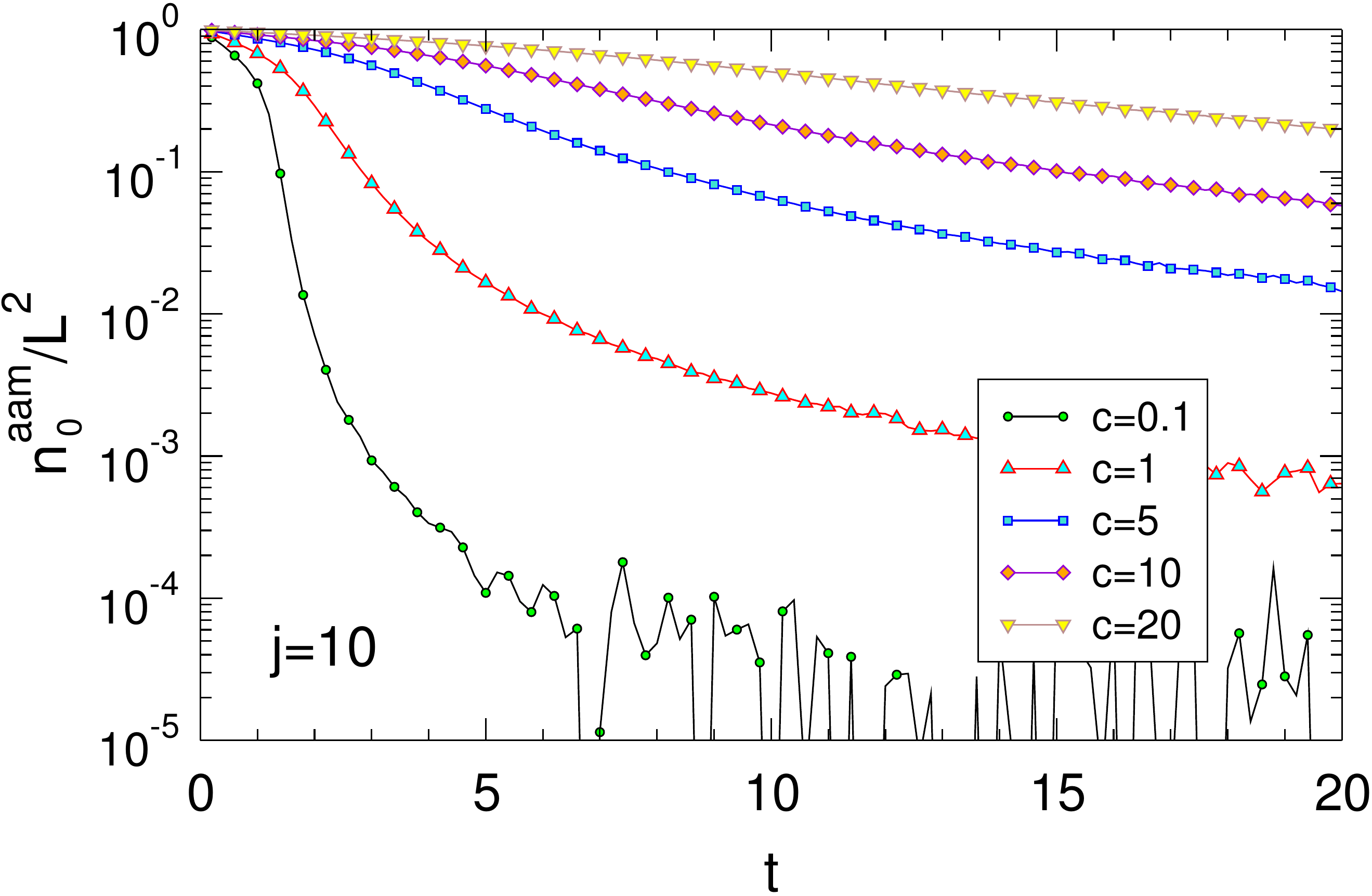}
\caption{ Coupling coherence $n^{aam}_0$ (Eq.~\eqref{coupling_coherence}) extrapolated to the thermodynamic limit $L\to\infty$ for $j=10$ and various $c$. 
It always takes a finite value, at all temperature, showing that $\langle e^{i(2\phi_{\bm r}^a-\phi_{\bm r}^m)}\rangle\propto \sqrt{n^{aam}_0}$  is finite at all temperatures. 
}
\label{Linfinite_Saam}
\end{center}
\end{figure}

\section{Ordering of $\cos(2\phi^a_{\bf r}-\phi^m_{\bf r})$ at all temperatures}
\label{a.naam}
The fact that the angular variable $\Omega_{\bf r}= 2\phi^a_{\bf r}-\phi^m_{\bf r}$ orders at all temperature for any finite couplings $C$ is most easily seen in studying its susceptibility
\begin{equation}
n^{aam}_0= \frac{1}{L^2} \left \langle  \sum_{{\bm r},{\bm r'}}  \cos(\Omega_{{\bm r}{\bm r}'})  \right \rangle~,
\label{coupling_coherence}
\end{equation}
where    $\Omega_{{\bm r}{\bm r}'} = \Omega_{\bm r} - \Omega_{{\bm r}'}$.  Because of $C$, we expect  the connected  correlation function of the field $\Omega_{\bm r}$
\begin{equation}
G_\Omega(r)= {\rm Re} \left [ \langle e^{i\Omega_{\bm r}}e^{-i\Omega_0}\rangle - |\langle e^{i\Omega_{\bm r}}\rangle |^2 \right ]~,
\end{equation}
to be exponentially decreasing at long distance, implying that 
\begin{equation}
\frac{n^{aam}_0}{L^2}=\frac{\alpha}{L^2}+ |\langle e^{i\Omega_{\bm r}}\rangle|^2,
\label{e.naam}
\end{equation}
 where $\alpha=\sum_{\bm r} {\rm Re}\, G_\Omega(r)$ is a constant independent of $L$ for sufficiently large sizes. Thus  $n^{aam}_0/L^2>0$ in the thermodynamic limit implies that $\langle e^{i\Omega_{\bm r}}\rangle$ is finite.

 Fig.~\ref{Linfinite_Saam} shows the extrapolation of $n^{aam}_0/L^2$ to the thermodynamic limit  ($L$$\to$$\infty$) for various values of $c$ as a function of the temperature. We have used a fit function of the form $n^{aam}_0/L^2=\alpha/L^2+\beta/L+\gamma$, which fits very well the data with $\beta\approx 0$, consistently with Eq.~\eqref{e.naam}.  We clearly see that it is always finite at all temperature, implying that $\langle e^{i(2\phi_{\bm r}^a-\phi_{\bm r}^m)}\rangle\neq 0$, that is, the two phases are always locked asymptotically.

\end{document}